\documentclass[a4paper, twoside, reqno, 12pt, dvips]{amsart}
\RequirePackage{fix-cm} 

\usepackage{fixltx2e}     

\usepackage{amssymb}
\usepackage{amsmath}
\usepackage{mathrsfs, dsfont}

\usepackage{a4wide}

\usepackage{acronym}

\usepackage{indentfirst}
\usepackage{graphicx}
\usepackage{psfrag}
\usepackage{epstopdf}

\usepackage[usenames, dvipsnames, pdf]{pstricks}
\usepackage{epsfig}
\usepackage{pst-grad} 
\usepackage{pst-plot} 

\usepackage{subfigure}

\usepackage{longtable}

\usepackage[french,english]{babel}
\usepackage[latin1]{inputenc}

\usepackage{color}
\definecolor{oneblue}{rgb}{0,0.0,0.75}
\usepackage[colorlinks,
            urlcolor=oneblue,
            linkcolor=oneblue,
            citecolor=oneblue,
            bookmarksopen=false,
            pagebackref]{hyperref}

\numberwithin{equation}{section}

\begin{document}

\title[On the modelling of tsunami waves]{On the modelling of tsunami generation and tsunami inundation}

\author[F. Dias]{Fr\'ed\'eric Dias$^*$}
\address{School of Mathematical Sciences, University College Dublin, Belfield, Dublin 4, Ireland \and CMLA, UMR 8536 CNRS, Ecole Normale Sup\'erieure de Cachan, Cachan, France}
\email{Frederic.Dias@ucd.ie}
\thanks{$^*$ Corresponding author}

\author[D. Dutykh]{Denys Dutykh}
\address{School of Mathematical Sciences, University College Dublin, Belfield, Dublin 4, Ireland \and LAMA, UMR 5127 CNRS, Universit\'e de Savoie, Campus Scientifique, 73376 Le Bourget-du-Lac Cedex, France}
\email{Denys.Dutykh@univ-savoie.fr}
\urladdr{http://www.lama.univ-savoie.fr/~dutykh/}

\author[L. O'Brien]{Laura O'Brien}
\address{School of Mathematical Sciences, University College Dublin, Belfield, Dublin 4, Ireland}
\email{loliwarm@gmail.com}

\author[E. Renzi]{Emiliano Renzi}
\address{School of Mathematical Sciences, University College Dublin, Belfield, Dublin 4, Ireland}
\email{Emiliano.Renzi@ucd.ie}

\author[T. Stefanakis]{Themistoklis Stefanakis}
\address{CMLA, UMR 8536 CNRS, Ecole Normale Sup\'erieure de Cachan, Cachan, France and School of Mathematical Sciences \and University College Dublin, Belfield, Dublin 4, Ireland}
\email{Stefanakis.Themistoklis@gmail.com}

\begin{abstract}
While the propagation of tsunamis is well understood and well simulated by numerical models, there are still a number of unanswered questions related to the generation of tsunamis or the subsequent inundation. We review some of the basic generation mechanisms as well as their simulation. In particular, we present a simple and computationally inexpensive model that describes the seabed displacement during an underwater earthquake. This model is based on the finite fault solution for the slip distribution under some assumptions on the kinematics of the rupturing process. We also consider an unusual source for tsunami generation: the sinking of a cruise ship. Then we review some aspects of tsunami run-up. In particular, we explain why the first wave of a tsunami is sometimes less devastating than the subsequent waves. A resonance effect can boost the waves that come later. We also look at a particular feature of the 11 March 2011 tsunami in Japan -- the formation of macro-scale vortices -- and show that these macro-scale vortices can be captured by the nonlinear shallow water equations.
\end{abstract}
  

\maketitle

\tableofcontents

\section{Introduction}

the water surface is perturbed. There are three main types of disturbances: underwater earthquakes concentrated in zones where there is slipping or subduction of tectonic plates, submarine landslides which are often but not always triggered by earthquakes, and sudden earth surface movements adjacent to the ocean (volcanoes, rock falls, sub-aerial landslides, ship sinking). In the generation of tsunamis by earthquakes the key point is to predict the displacement of the sea bottom, and then to understand the energy transfer to the water column. The generation of tsunamis by submarine landslides is even more challenging for various reasons: lack of data, coupling between fluid and solid motions, longer duration, more physical parameters. The modeling of sudden earth surface movements adjacent to the ocean has not been much studied, the reason being that such movements are quite rare.

The propagation of tsunamis is now well understood and operational codes can easily propagate tsunamis across a whole ocean (see for example the various simulations of the megatsunamis of 2004 in the Indian Ocean and of 2011 in the Pacific Ocean).

Wave run-up (maximum vertical extent of wave uprush on a beach above still water level) and wave inundation (maximum horizontal extent) have been studied during the last sixty years, but continue to be a challenging problem. A high level of mesh refinement as well as high resolution bathymetric and topographic data are required to describe local wave run-up. 

Our group has developed a novel tool for tsunami wave modelling. This tool has the potential of being used for operational purposes: indeed, the numerical code VOLNA is able to handle the complete life-cycle of a tsunami (generation, propagation and run-up along the coast). The algorithm works on unstructured triangular meshes and thus can be run in arbitrary complex domains. A detailed description of the finite volume scheme implemented in the code as well as the numerical treatment of the wet/dry transition can be found in \cite{Dutykh2009a}.

In Section \ref{sec:gen}, we review some of the basic generation mechanisms as well as their simulation. In particular, we present a simple and computationally inexpensive model that describes the seabed displacement during an underwater earthquake. This model is based on the finite fault solution for the slip distribution under some assumptions on the kinematics of the rupturing process. We also consider an unusual source for tsunami generation: the sinking of a cruise ship. In Section \ref{sec:inun}, we review some aspects of tsunami run-up. In particular, we explain why the first wave of a tsunami is sometimes less devastating than the subsequent waves. A resonance effect can boost the waves that come later. We also look at a particular feature of the 11 March 2011 tsunami in Japan -- the formation of macro-scale vortices -- and show that these macro-scale vortices can be captured by the nonlinear shallow water equations.

\section{Tsunami generation}\label{sec:gen}

The modelling of tsunami generation was initiated in the early 1960's by the prominent work of \textsc{Kajiura} \cite{kajiura}, who proposed the static approach which is still widely used by the tsunami wave modelling community: the static sea bed displacement is translated towards the free surface as an initial condition. The most classical solution for the co-seismic sea bed displacements is the celebrated Okada solution \cite{Okada85}. The kinematics of earthquakes is relatively well understood. The maximum bottom deformation is achieved during a finite time known as the rise time. For example the rise time was $8$ seconds for the July 17, 2006 Java event
simulated by \textsc{Dutykh} \emph{et al}. \cite{Dutykh2010a}. In the next subsection, we will show how kinematics can be taken into account in tsunami generation.

However, modelling tsunamis generated by landslides is far more complicated. Firstly, the kinematics of underwater landslides is not well understood. Also the time scale over which they occur is longer than for an earthquake, so simply transferring the sea bed deformation directly to the free surface does not accurately model the landslide. For one-dimensional landslides, \textsc{Liu} \emph{et al}. \cite{Liu2003} showed that the free-surface elevation is significantly different from the bottom motion by using an anlytical technique developed by \textsc{Tuck} \& \textsc{Hwang} \cite{Tuck1972}. \textsc{Sammarco} \& \textsc{Renzi} \cite{Sammarco2008} extended the results to two-dimensional landslides (see also \cite{Renzi2010, Renzi2012}). \textsc{Sarri} \emph{et al}. \cite{Sarri2012} used the \textsc{Sammarco} \& \textsc{Renzi} model to build a statistical emulator.

Mass movements on dry land can be put into a number of different categories. Rotational slides, translational slides, block slides, falls, topples, debris flows, debris avalanches, earthflows, creep and lateral spreads each have their own characteristic kinematics of motion. However, when considering landslides that are in contact with water, our knowledge of the kinematics in this environment is lacking. This is due to lack of data, in particular lack of bathymetry data prior to large tsunamigenic events.

A recent review of submarine mass movements \cite{Locat2009} shows that they can consist of soil and rock, and similar to dry land movements they can take the form of slides, spreads, flows, topples or falls, but in addition they can develop into turbidity currents. They can move up to to $50$ km/h, reach distances over $1000$ km and volumes can be enormous, the largest known being the Storegga  slide at approximately $2500$ km$^3$. Underwater mass movements pose a threat to coastal communities and infrastructures both onshore and offshore. The main triggers are seismic shaking, overloading gas hydrate dissolution and excess pore pressure, wave loading, erosion and human activities such as coastal construction. The 1998 Papua New Guinea tsunami generated a renewed interest in tsunamigenic landslides.

Although generally landslide tsunamis occur over a much smaller scale than earthquakes ($O(1 \mathrm{km})$ vs $O(100 \mathrm{km})$), they can cause very large run-up values. \textsc{Gutenberg} \cite{Gutenberg1939} was one of the first to suggest that tsunamis can be caused by submarine landslides. He goes so far as to say that they can be considered as one of the chief causes of tsunamis. He reported that co-seismic landslides possibly triggered large waves in Ceram in 1899, Assam in 1897 and caused cable breaks in Greece in the 19th century. More recently, in 1979, a part of the Nice harbour extension slumped into the Mediterranean and was followed by a small tsunami \cite{Assier-Rzadkieaicz2000}.

The Storegga slide is one of the largest known submarine landslides ($2500-3500$ km$^3$) and occurred off the west coast of Norway generating a huge tsunami 8200 years ago \cite{Bryn2005}. In 1929 an earthquake at the edge of Grand Banks, Canada, triggered a large submarine slope failure ($200$ km$^3$) generating a tsunami with run-up heights up to $13$ m that propagated as far as Portugal and the Azores Islands \cite{Fine2005}. A tsunami generated by a $0.03$ km$^3$ of rock falling from $914$ m into Lituya Bay, Alaska, in 1958 is dubbed the world's largest tsunami at $524$ $m$. \textsc{Plafker} \cite{Plafker1965} points out numerous major landslides that were triggered during the 1964 Alaska Earthquake. 

There is debate about the triggering of the 1946 Aleutian tsunami that destroyed Scotch Cap's lighthouse. \textsc{Freyer} et al. \cite{Fryer2004} argue the possible involvement of a submarine slump that caused nearsource damage. Similarly, landslides have been suggested as a possible cause of amplification of the 1992 Flores Island tsunami \cite{Imamura1995}. \textsc{Ma} \emph{et al}.\cite{Ma1999} demonstrate large scale slumping combined with faulting as a good model for the 1975 Kalapana, Hawaii tsunami.

Seafloor mapping has begun to open up the research of submarine landslides. Evidence of several underwater slides in the St. Lawrence estuary, Canada, have been identified through seafloor mapping \cite{PoncetCanada2010}.

The scale of the 1998 Papua New Guinea tsunami was a ``wake-up call'' for tsunami scientists. Although there was much controversy initially over the main trigger of the tsunami, the magnitude $7.1$ earthquake was relatively small compared to the $10-15$ m tsunami that devastated the coast. \textsc{Synolakis} \emph{et al}. \cite{Synolakis2002} presented high resolution bathymetric data combined with hydrodynamic modelling indicating a large underwater slump in the area. Subsequent research of this event has contributed to new and improved models of tsunamis due to submarine mass failure and it is now accepted by most scientists that the source of this event was a submarine slump \cite{Tappin2008}.

There is also strong evidence that Island volcanoes such as Stromboli, Italy \cite{Tinti2000a}, Ritter Island, Papau New Guinea \cite{Ward2003} and the Hawaiian Islands \cite{Moore1994} have experienced lateral flank collapses in the distant past which potentially could have caused large scale tsunamis. On the 30\up{th} December 2002 part of the western flank of the Stromboli volcano slid into the sea initiating two tsunamis seven minutes apart \cite{Locat2009}. This gave \textsc{Tinti} \emph{et al}. \cite{Tinti2005} and \textsc{Chiocci} \emph{et al}. \cite{Chiocci2008} the unique opportunity to observe and model both the submarine and subaerial morphological changes thanks to a previous multibeam survey of the area. Recently, scientists have become concerned that a section of the Cumbre Vieja volcano, La Palma in the Canary Islands, may experience failure and generate a megatsunami \cite{Day1999, Ward2001}.

Landslides have the potential to cause more destructive tsunamis than originally thought. They may be seismic triggered landslides amplifying waves originally generated from earthquakes, large scale collapsing of volcanoes into the sea, or rock falls from mountainous regions into restricted bays. In Section \ref{sec:costa}, we will consider a special case of tsunami, generated by the sinking of a cruise ship.

\subsection{Finite fault}

\begin{figure}%
\centering
\subfigure[$t=20$ $s$]%
{\includegraphics[width=0.45\textwidth]{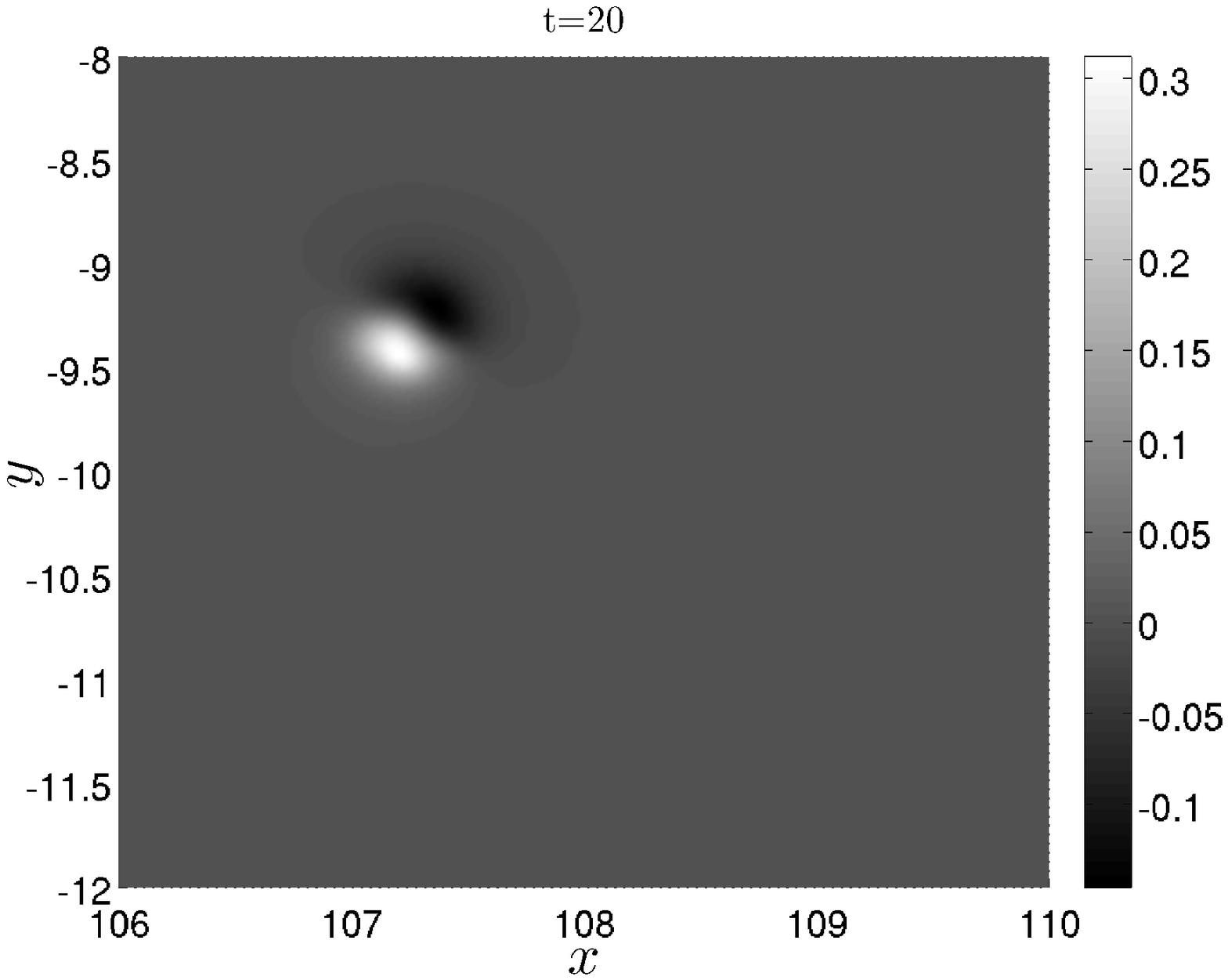}}
\subfigure[$t=50$ $s$]%
{\includegraphics[width=0.45\textwidth]{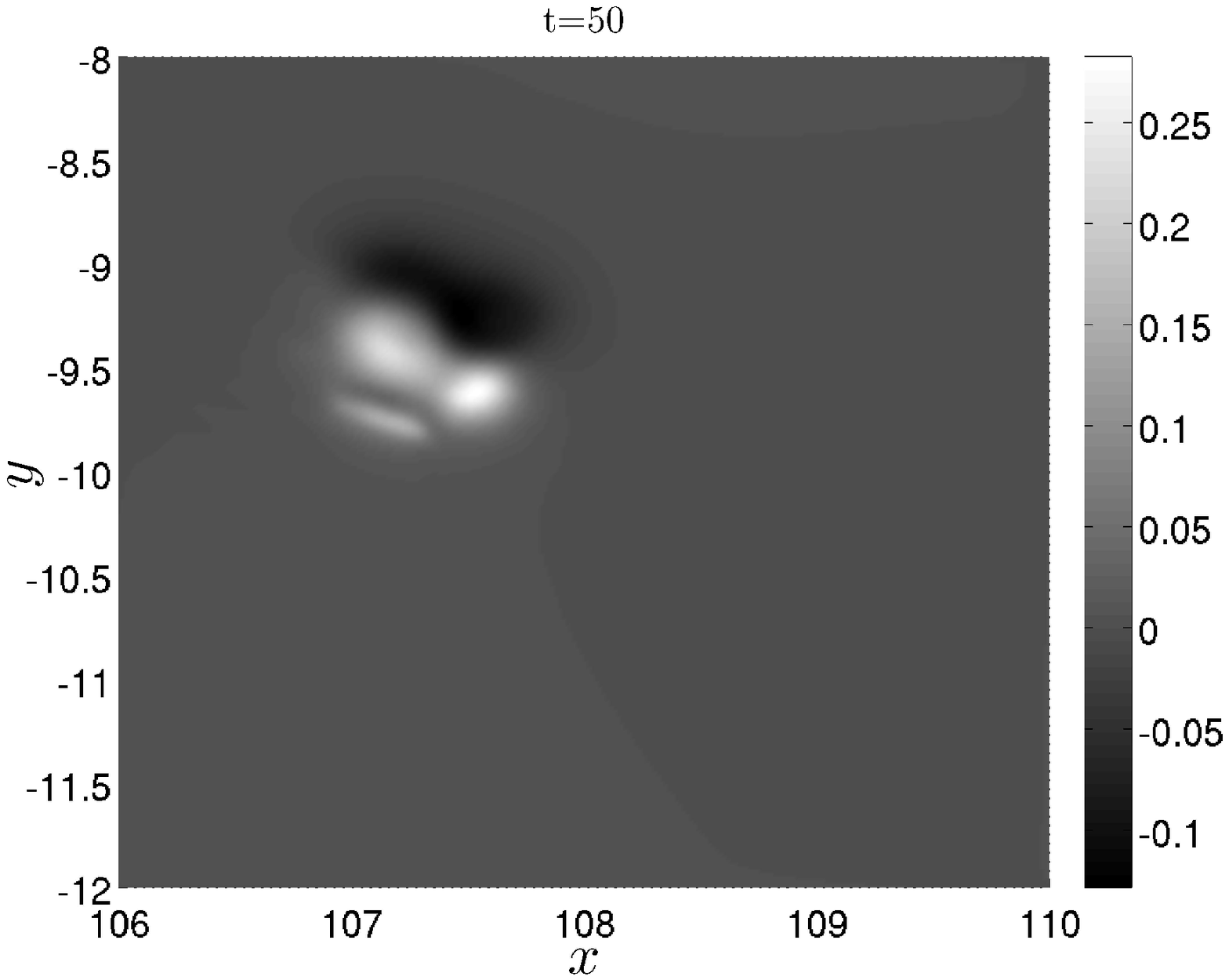}}
\subfigure[$t=80$ $s$]%
{\includegraphics[width=0.45\textwidth]{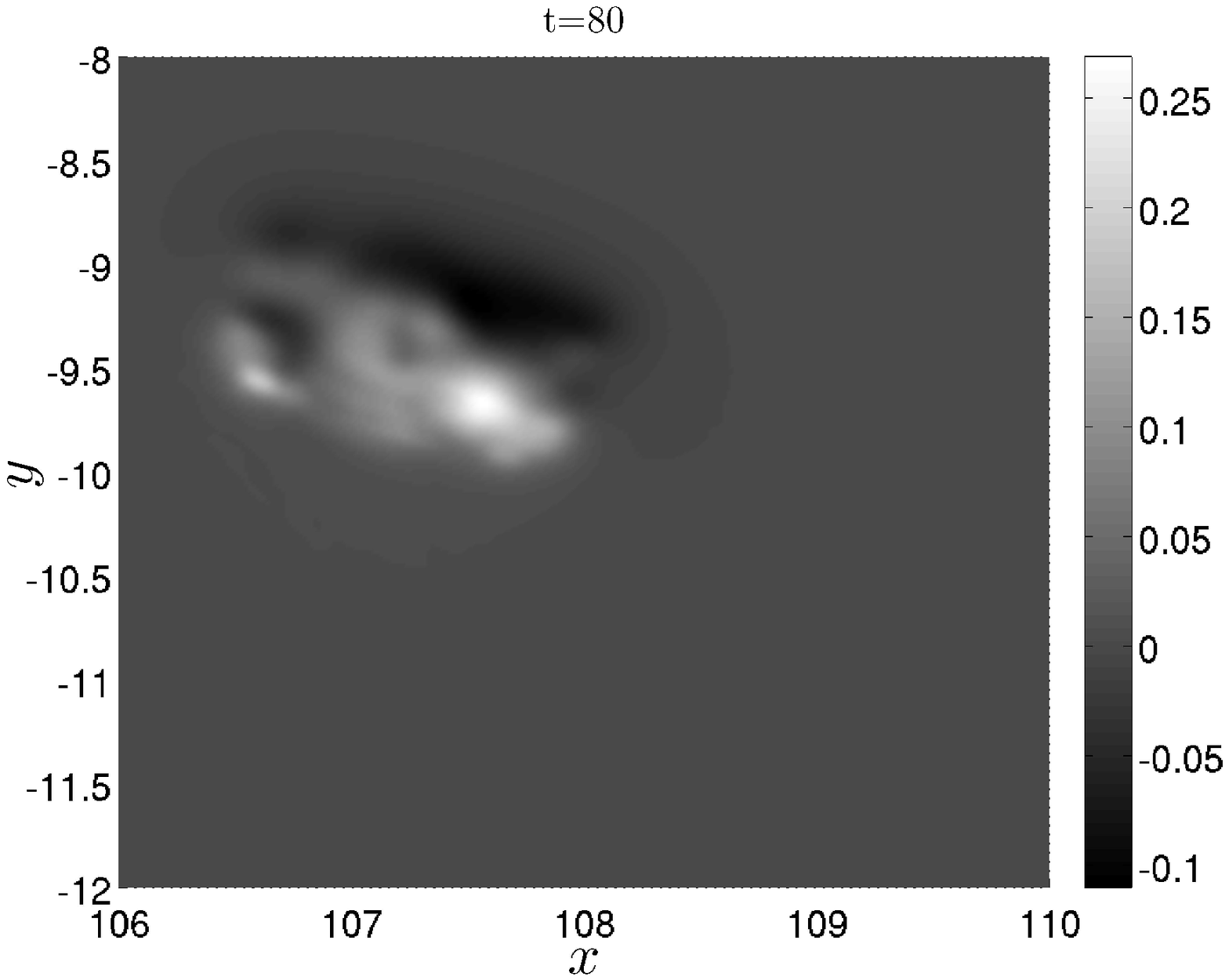}}
\subfigure[$t=140$ $s$]%
{\includegraphics[width=0.45\textwidth]{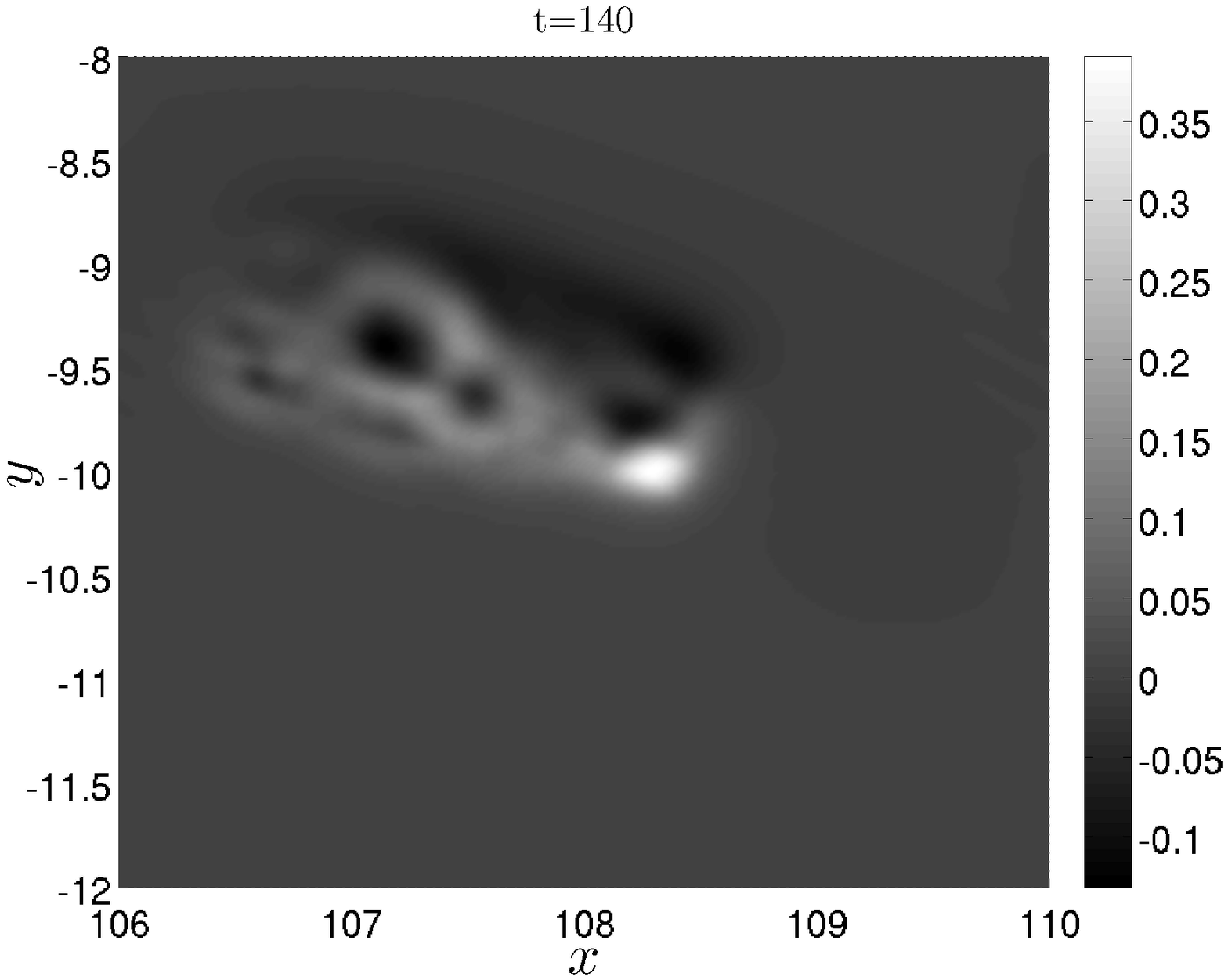}}
\subfigure[$t=200$ $s$]%
{\includegraphics[width=0.45\textwidth]{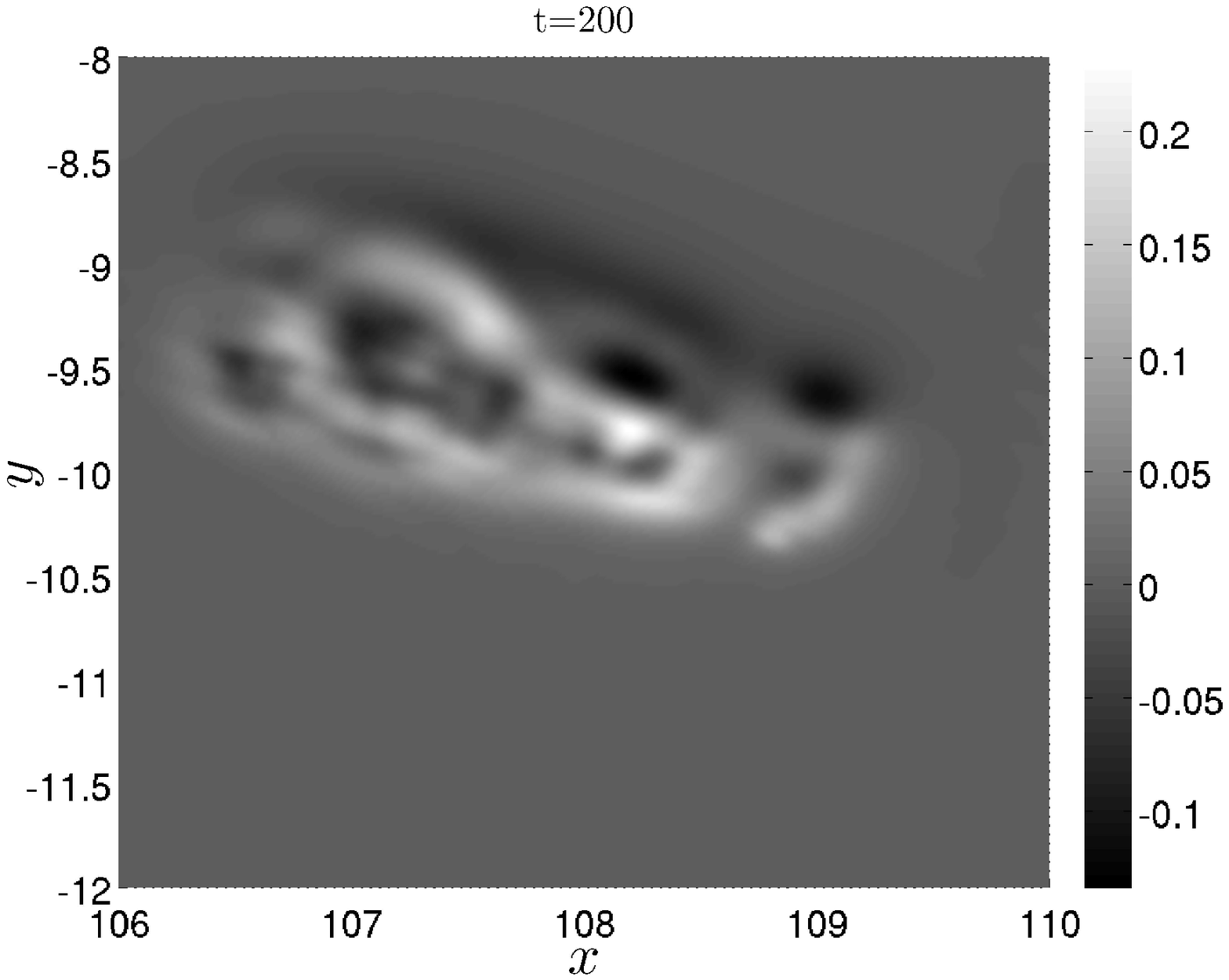}}
\subfigure[$t=250$ $s$]%
{\includegraphics[width=0.45\textwidth]{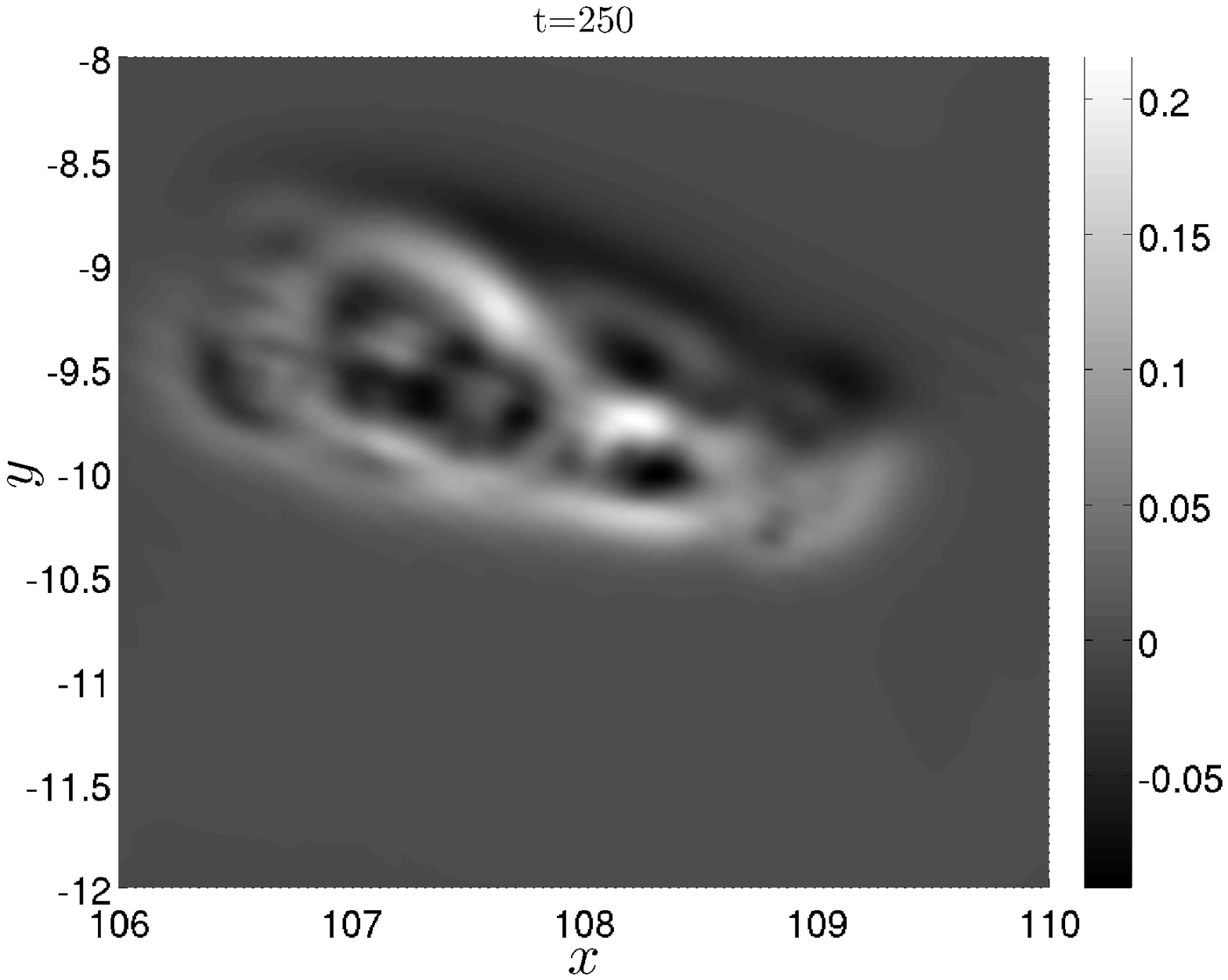}}
\caption{Generation of the July 17, 2006 tsunami in Java. Snapshots of the free surface elevation computed with a water wave model. The waves are generated by dynamic co-seismic bottom displacements reconstructed using the corresponding finite fault solution.}%
\label{fig:soleuler}%
\end{figure}

Here we review some recent advances in seismology and show how to reconstruct better co-seismic displacements of a tsunamigenic earthquake. More precisely, we use the so-called finite fault solution developed by Ji and his collaborators \cite{Ji2002}, based on static and seismic data inversion. This solution provides multiple fault segments of variable local slip, rake angle and several other parameters. By applying Okada's solution to each subfault, the sea bed displacement is reconstructed with higher resolution. Since Okada's solution consists of relatively simple closed-form analytical expressions, all computations can be done efficiently enough so that they can be used in a real-time Tsunami Warning System. Seabed displacements are then coupled with a water wave model. Further details can be found in \cite{Dutykh2010a}. This approach was recently extended to include horizontal displacements as well \cite{Dutykh2010d}. Figure \ref{fig:soleuler} shows snapshots of the free surface elevation for the July 17, 2006 tsunami generation. The $x-$axis is the longitude while the $y-$axis is the latitude. The times are in seconds. The water elevation is in meters.

\subsection{Costa Concordia}\label{sec:costa}

\begin{figure}
\begin{center}
\includegraphics[width=0.9\textwidth]{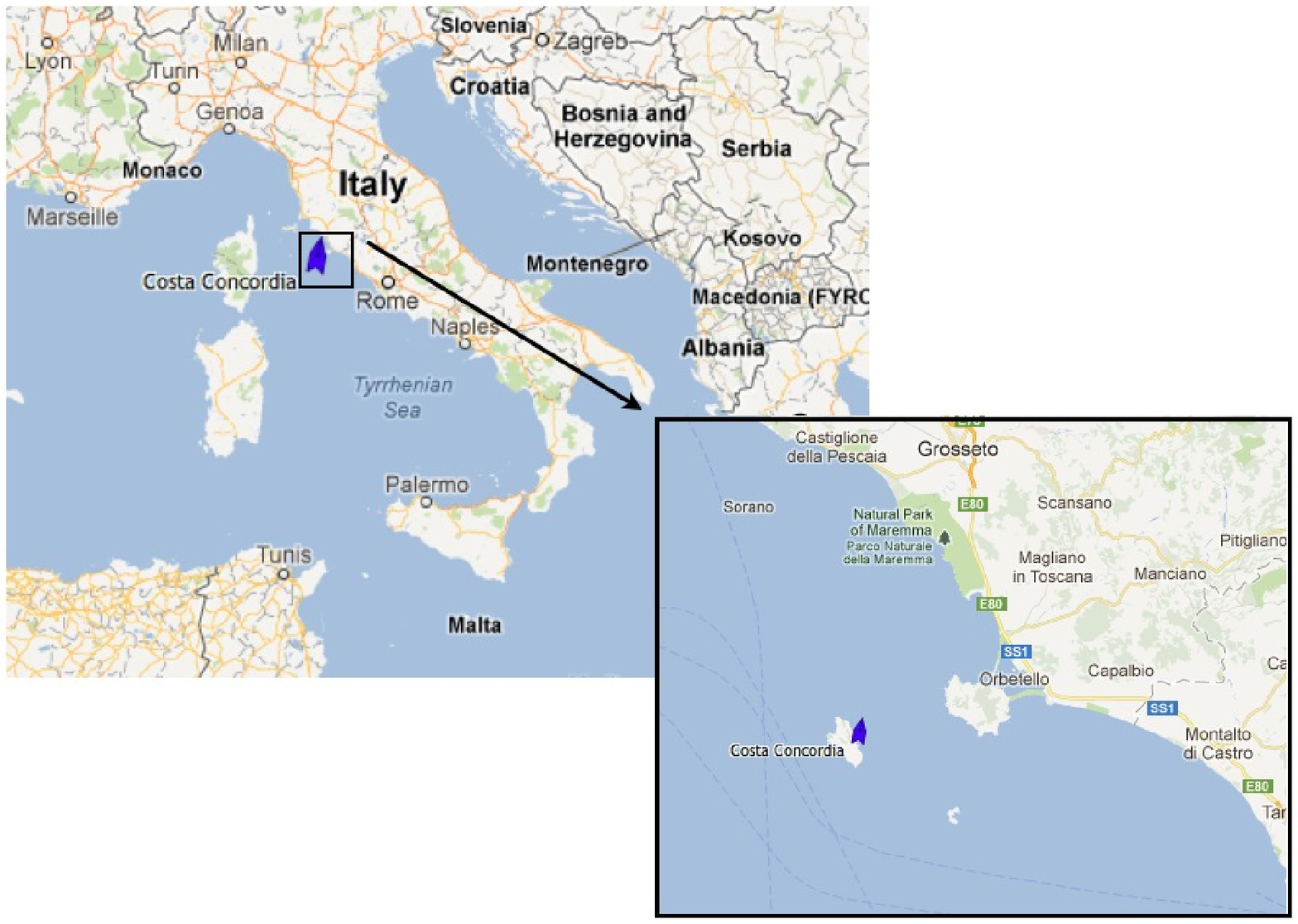}
\caption{Location of the Costa Concordia at time of submission from \url{http://www.marinetraffic.com} (17/05/12)}
\label{Google_maps}
\end{center}
\end{figure}

\begin{figure}
\begin{center}
\includegraphics[width=0.7\textwidth]{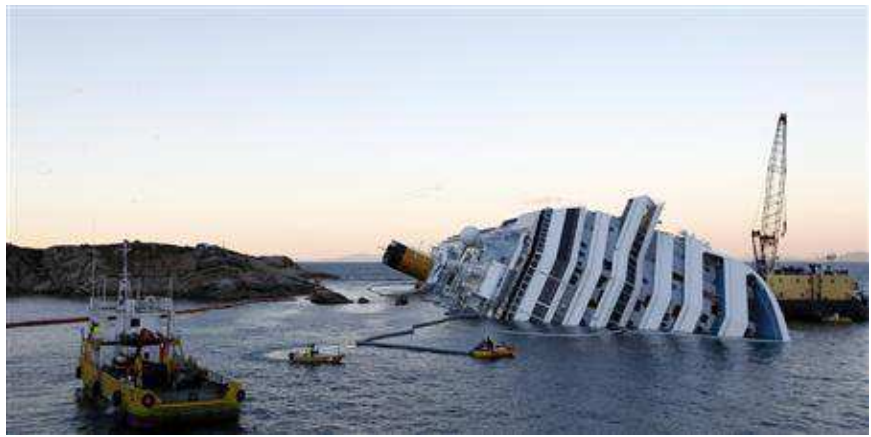}
\caption{The Costa Concordia {\it uk.reuters.com} (retrieved 17/05/12)}
\label{Costa}
\end{center}
\end{figure}

\begin{figure}
\begin{center}
\includegraphics[width=0.7\textwidth]{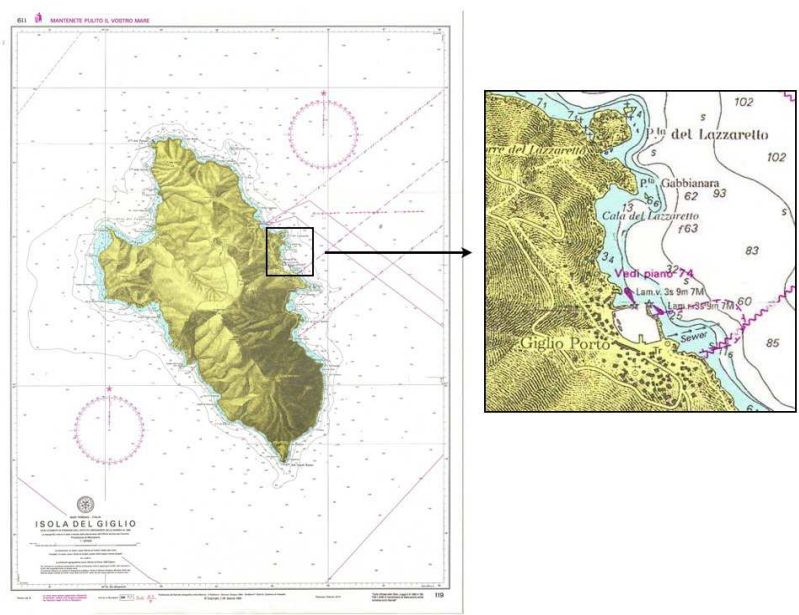}
\caption{Bathymetric chart from the Italian Hydrographic Institute. The Costa Concordia is near Gabbianara (north of Giglio Porto on the map)}
\label{Chart}
\end{center}
\end{figure}

\begin{figure}
\begin{center}
\includegraphics[width=0.7\textwidth]{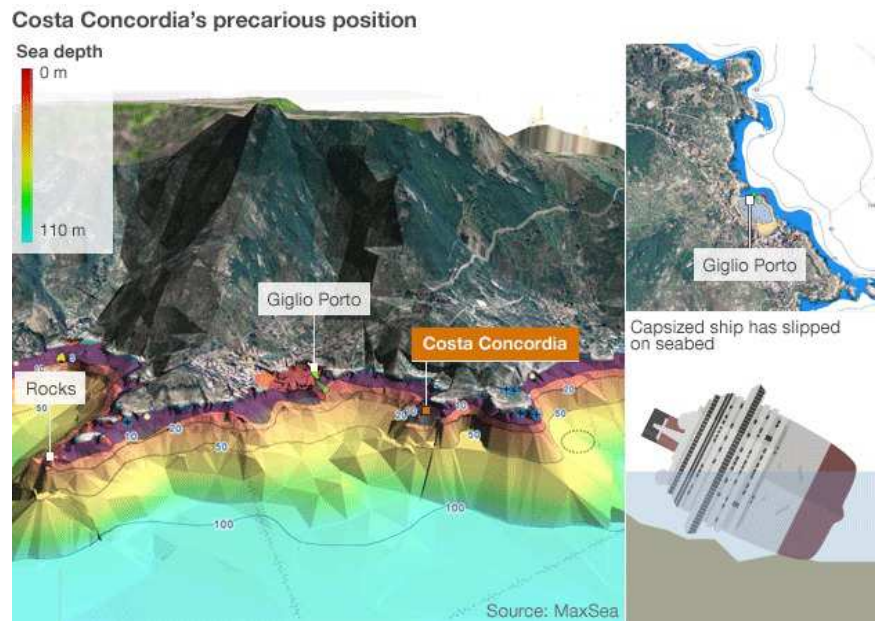}
\caption{Costa Concordia position from BBC news \url{http://www.bbc.co.uk/news/world-europe-16563562} (retrieved 17/05/12)}
\label{BBC}
\end{center}
\end{figure}

This is the first application of the VOLNA code presented in this paper. A second one will be presented below. The Costa Concordia has been balancing nearly horizontally on its side since she partially sank on the night of 13 January 2012 on rocks beside the island of Giglio, Italy (see Figures \ref{Google_maps} and \ref{Costa}). The local bathymetry in the area contains a steep drop from approximately $20$ m depth to over $100$ m (see Figures \ref{Chart} and \ref{BBC}). Before the ship is stabilised, it is possible that she may slide into deeper waters and create waves that would travel along the island and possibly head towards the Italian mainland. If this were to happen, could a tsunami be generated?

A similar incident occurred off the Greek island of Santorini in 2007 with the MS Sea Diamond when she ran aground on a volcanic reef. The ship was towed off the rocks and stabilised, but the ship sank some 15 hours after it initially struck rocks. The bathymetry in the area is a caldera (cauldron-like volcanic feature) and so the shore is almost vertical. The stern now sits in about $180$ m of water and the bow in about $60$m (information obtained through Wikipedia). It was feared the wreck would slide deeper into the caldera below.
 
We consider here estimates taken from \cite{Miloh1978}, Equation (\ref{eta_Murty_1}) and \cite{Okal2003}, Equation (\ref{gaussian_est}) for wave heights generated by displacements in the bathymetry:
\begin{equation}\label{eta_Murty_1}
\eta_1 = \left( 8 \alpha \frac{\rho_s} {\rho_w} \frac{ \ell h_Lv}{c} \right)^{1/2}
\end{equation}
and
\begin{multline}\label{gaussian_est}
 \eta_2(x,t) = \frac{h_L v^2}{2 c} \Bigl[  \frac{ \exp{(-k (x + ct)^2)} - \exp{(-k(\xi + c\tau)^2)}  }{ c + v  } + \\ \frac{ \exp{(-k (x -ct)^2)} - \exp{(-k(\xi - c\tau)^2)}  }{ c - v  } \Bigr],
\end{multline}
where $\xi = x-vT$ and $\tau = t -T$. In these expressions, $\eta$ denotes the free-surface elevation (maximum only for $\eta_1$ and space and time distribution for $\eta_2$). $\alpha$ is the transfer coefficient (only a portion of the landslide momentum is imparted on the water column), $\rho_s$ is the solid density, $\rho_w$ is the water density, $w$ and $\ell$ are the two horizontal dimensions of the landslide, $h_L$ is the height of the landslide, $v$ is the landslide velocity, $c$ is the long wave velocity ($\sqrt{gh}$) with $h$ the water depth, $T$ is the duration of the landslide and $k$ is a coefficient.

\begin{figure}
\begin{center}
\includegraphics[width=0.7\textwidth]{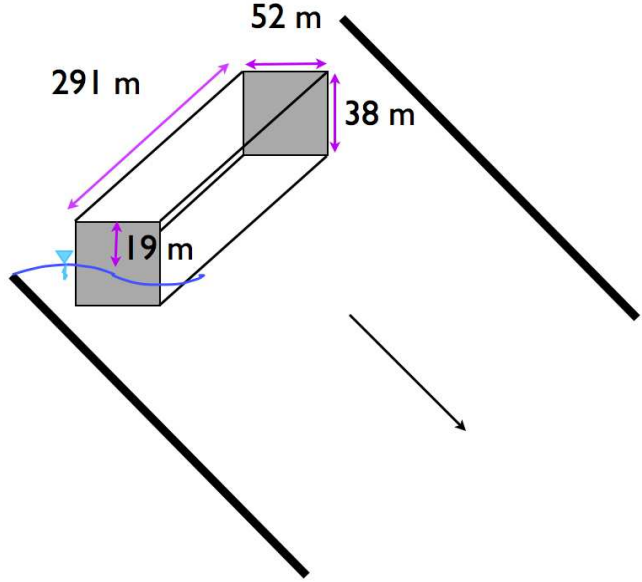}
\caption{Rough dimensions of the Costa Concordia}
\label{dim}
\end{center}
\end{figure}

The dimensions of the ship are $291$ x $52$ x $38$ m$^3$. Assuming that she is lying completely on her side, a rough geometry is shown in Figure \ref{dim}. The following assumptions are used: the width parallel to the shore is $w = 291$ m, the height is $h_L = 38$ m, the length is $\ell = 52$ m, the distance from the top of the ship to the still water line is $\approx 38/2 = 19$ m, the depth $h$ is $\approx 38/2 = 19$ m, and the slope of the shore is $\approx 80/400 = 0.2$. We also assume that $\rho_s = \rho_w = 1000$ kg/m$^3$, $\alpha = 0.01$ and $k = 18/\ell^2$. Then the only missing parameter is the velocity $v$, which of course is difficult to estimate. Some sources say that the Titanic sank at approximately $15$ m/s once she was completely submerged (information obtained through National Geographic). However, her stern lifted high into the air as the ship pivoted down into the water and she may have broken in two before she sank completely. Although the Costa Concordia is of the same scale as the Titanic (she is actually bigger) the Titanic was in open water and sank bow first, where as the Costa Concordia will have friction due to the rocks below her and is positioned lying on her side which may cause her to sink differently.

Applying the above parameters to Equation (\ref{eta_Murty_1}) from \cite{Miloh1978} gives
\begin{equation*}
\eta_1 = \left( (8)(0.01) (1) \frac{(52)(38)v}{\sqrt{(9.8)(19)}}\right) ^{1/2} = (11.6 v)^{1/2} \mbox{ m}.
\end{equation*}
Sinking speeds of $v = [0.001, 0.1, 1]$ m/s will give wave heights of $\eta = [0.108, 1.08, 3.41]$ m. Therefore, the speed that the ship goes down at is an important parameter to get right.

\textsc{Okal} \& \textsc{Synolakis} \cite{Okal2003} model the free surface propagation due to an underwater gaussian slump source by Equation (\ref{gaussian_est}). With the above parameters for the Costa Concordia together with a total travel time $T = 100$ s and a velocity $v = 1$ m/s, the maximum occurs at $x = 0$, $t = 0$. The free-surface elevation at this point is $\eta_2 (0,0) = 0.2052$ m. For $v = 0.1$ m/s, $\eta_2(0,0) = 0.002$ m and $v = 10$ m/s gives $\eta_2(0,0) = 44.08$ m. Again the velocity is an important parameter.

\begin{figure}
\begin{center}
\includegraphics[width=0.7\textwidth]{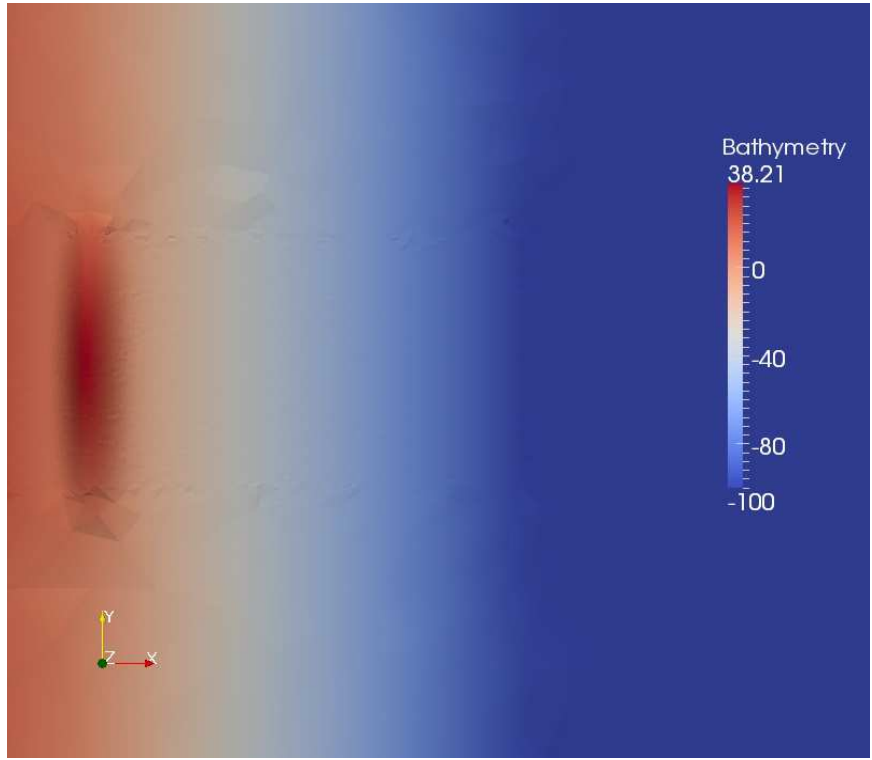}
\caption{Geometry used for the VOLNA simulation of the Costa Concordia on sloping sea floor, from above.}
\label{Costa_Bathymetry_close}
\end{center}
\end{figure}

Simulations of the sinking of the Costa Concordia were carried out using the VOLNA code. A gaussian shaped slide with the same volume as above ($291$ x $52$ x $38$ m$^3$) was used as a model of the ship and three simulations were carried out with sinking speeds $v = (0.5, 1, 10)$ m/s. The geometry is shown in Figure \ref{Costa_Bathymetry_close}.

\begin{figure}
\begin{center}
\includegraphics[width=0.7\textwidth]{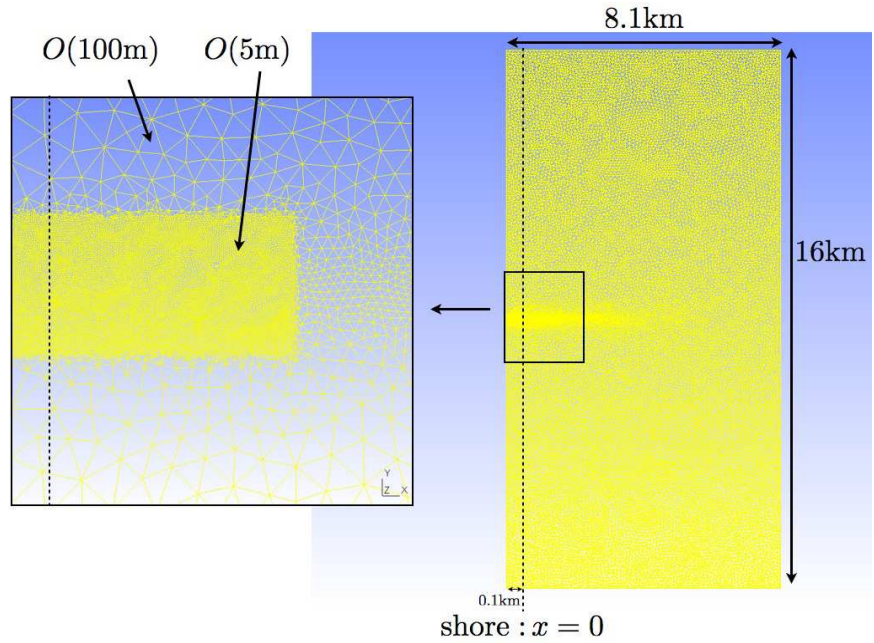}
\caption{Mesh used for the VOLNA simulation of the Costa Concordia.}
\label{costa_mesh}
\end{center}
\end{figure}

An unstructured triangular mesh was implemented over an area of $8.1$ x $16$ km$^2$ with a distance of $100$ m behind the shoreline $x = 0$. The control volumes were of the order of $100$ m, but refined to the order of $5$ m in a $600$ x $300$ m$^2$ region near the ship (see Figure \ref{costa_mesh}).

\begin{figure}
\begin{center}
\subfigure[$v = 0.5$ m/s]{
\includegraphics[width=0.31\textwidth]{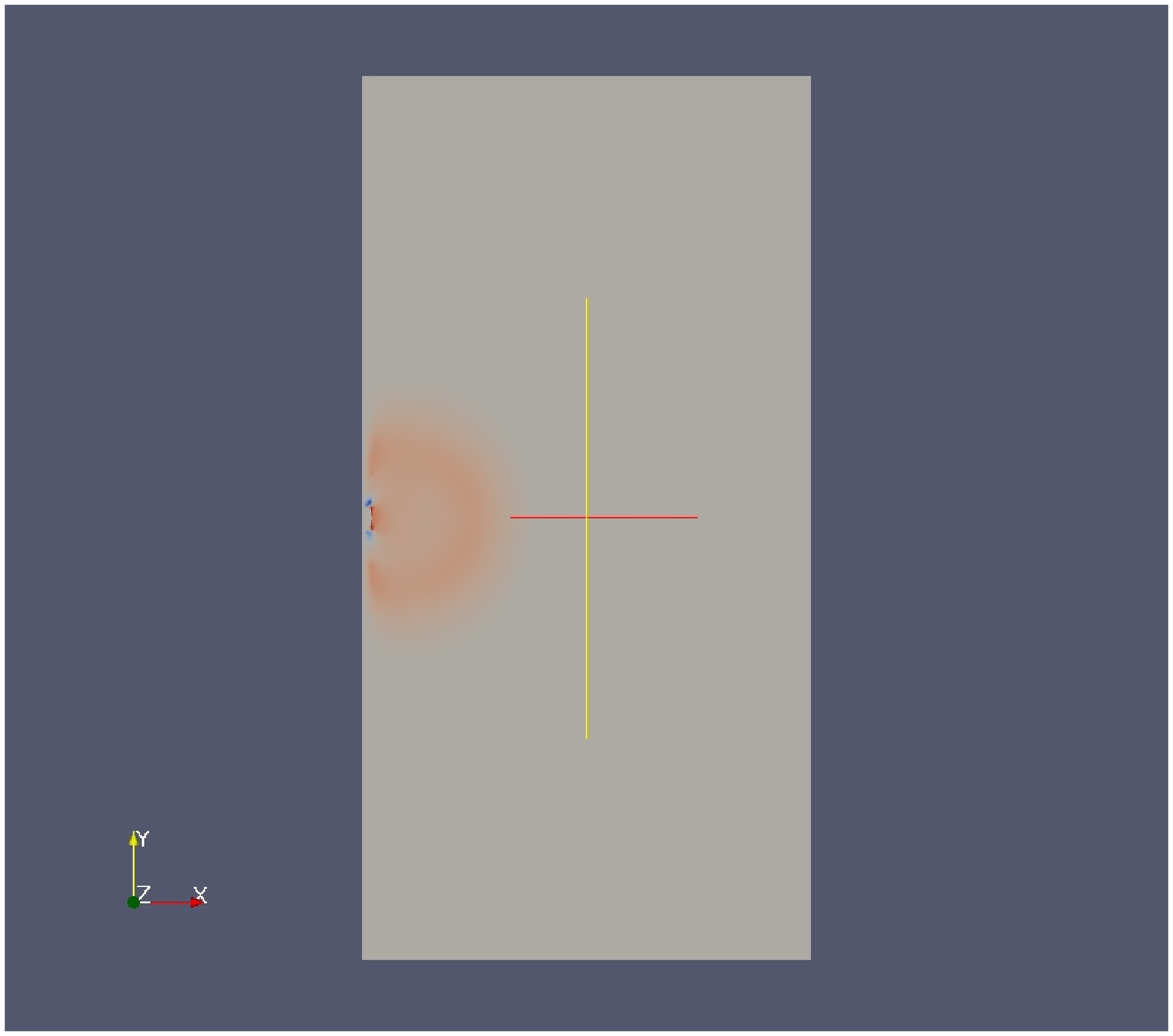} 
}
\subfigure[$v = 1$ m/s]{
\includegraphics[width=0.31\textwidth]{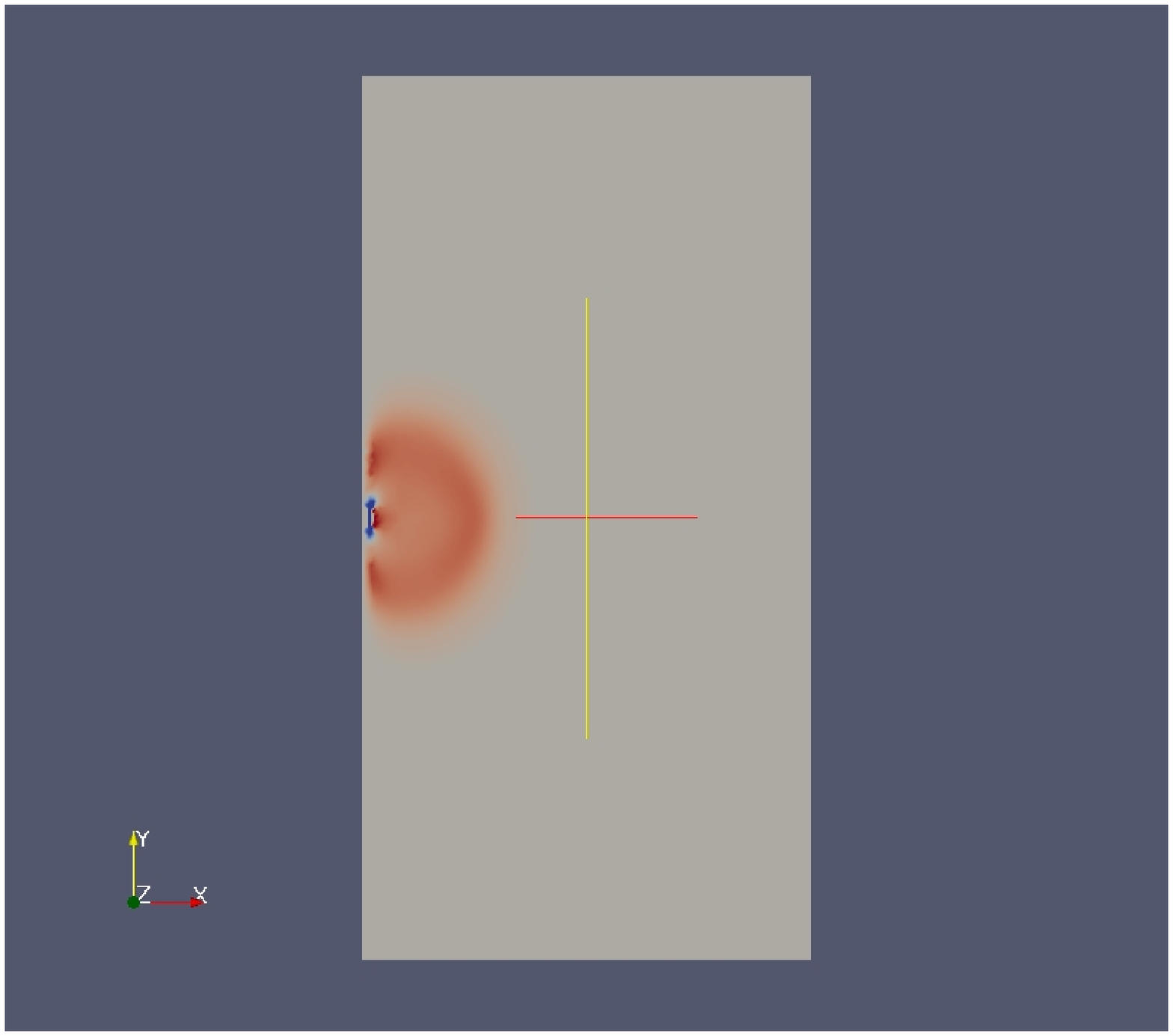} 
}
\subfigure[$v = 10$ m/s]{
\includegraphics[width=0.31\textwidth]{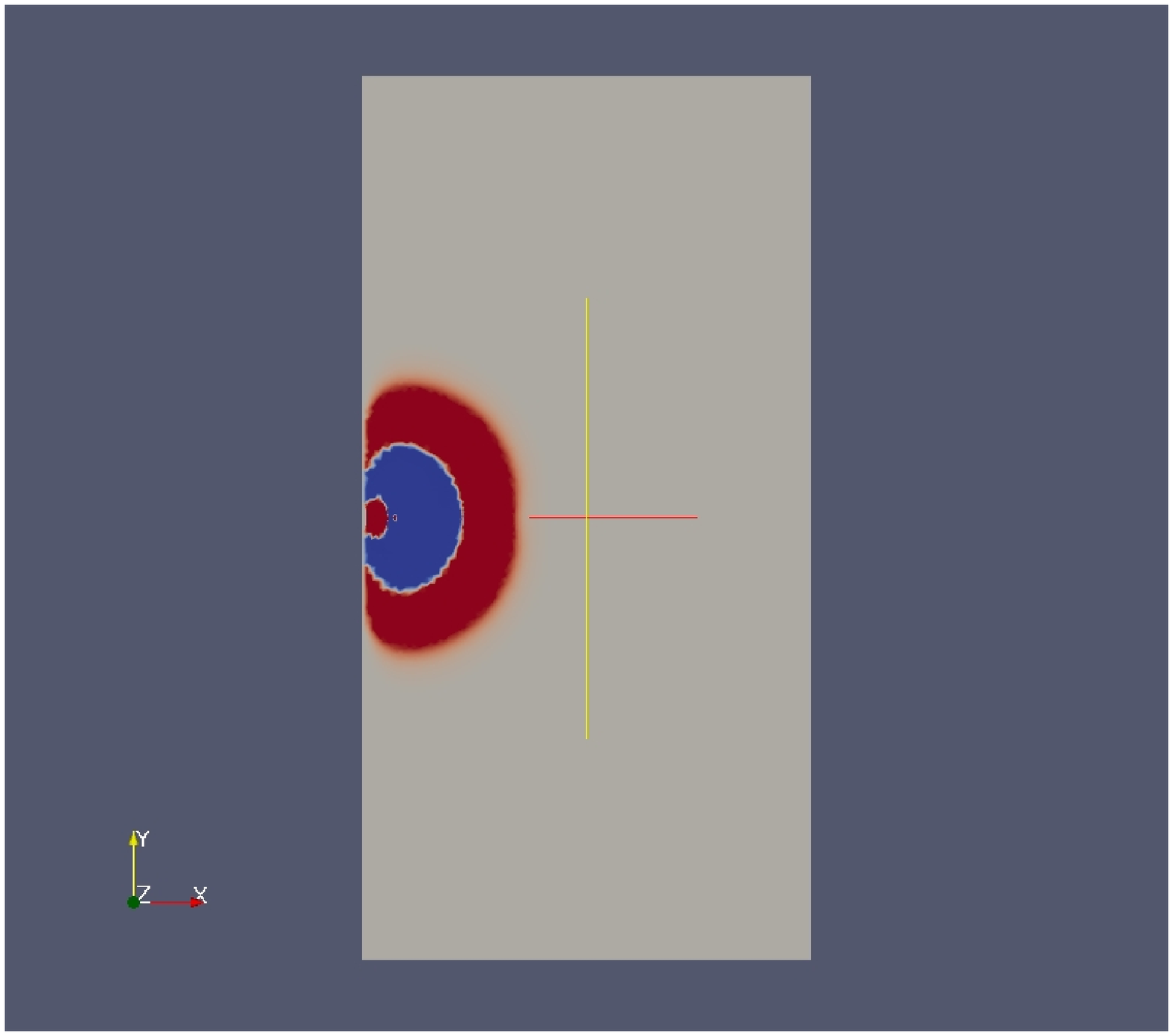} 
}
\end{center}
\caption{VOLNA simulations of the Costa Concordia sinking: free surface at $t = 75$ s.}
\label{costa_VOLNA_75s}
\end{figure}

\begin{figure}
\begin{center}
\subfigure[$v = 0.5$ m/s]{
\includegraphics[width=0.31\textwidth]{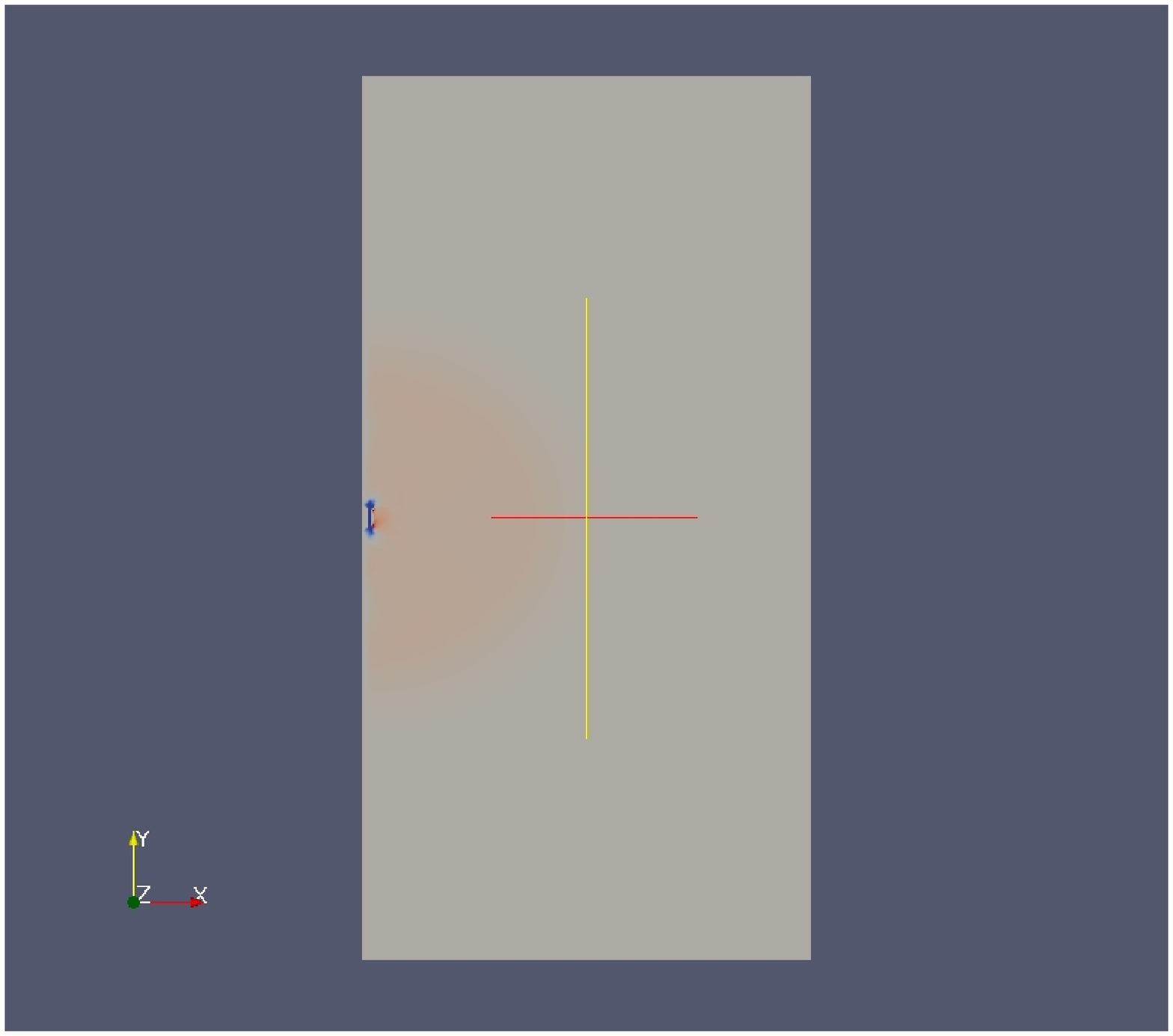}
}
\subfigure[$v = 1$ m/s]{
\includegraphics[width=0.31\textwidth]{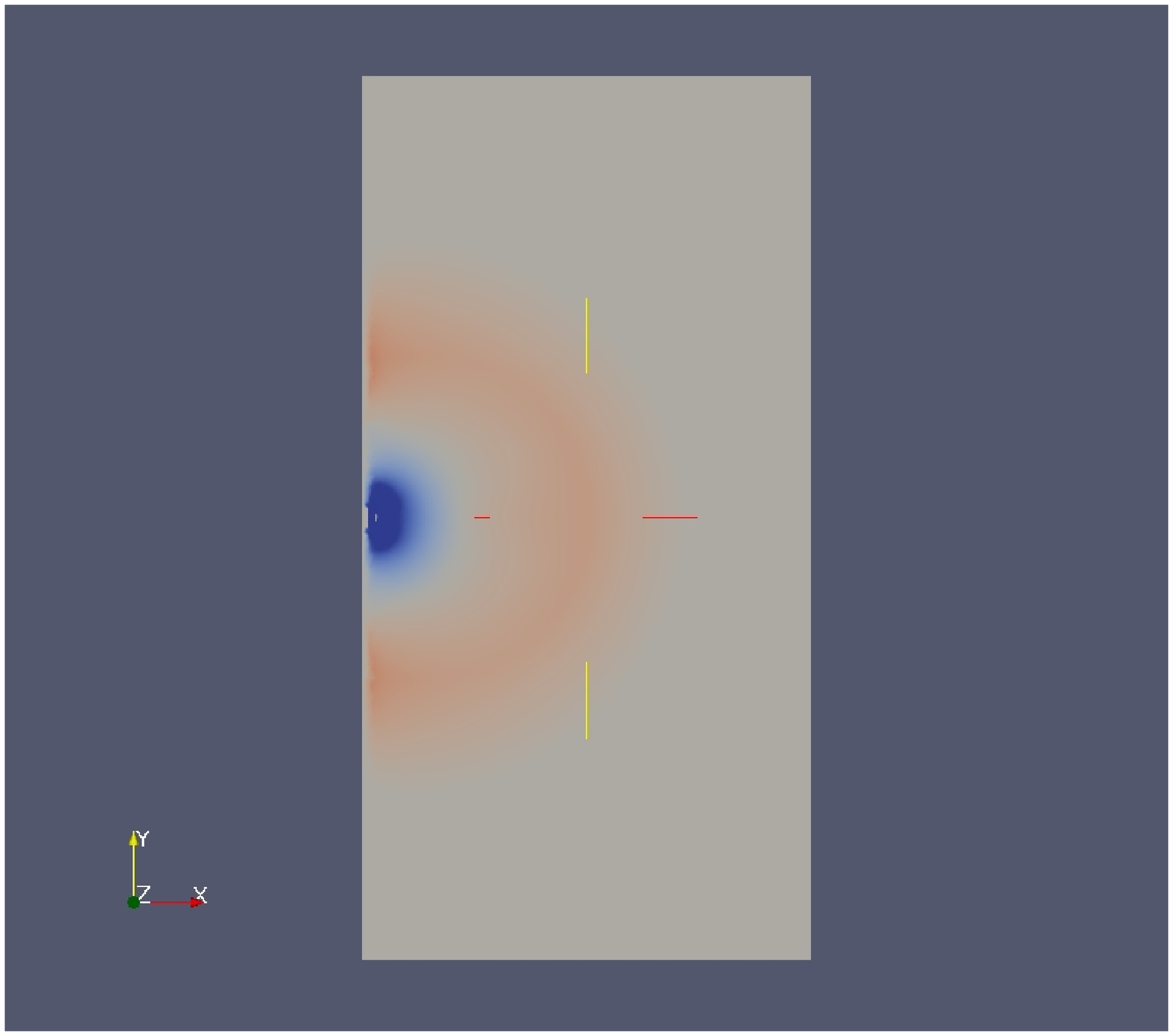}
}
\subfigure[$v = 10$ m/s]{
\includegraphics[width=0.31\textwidth]{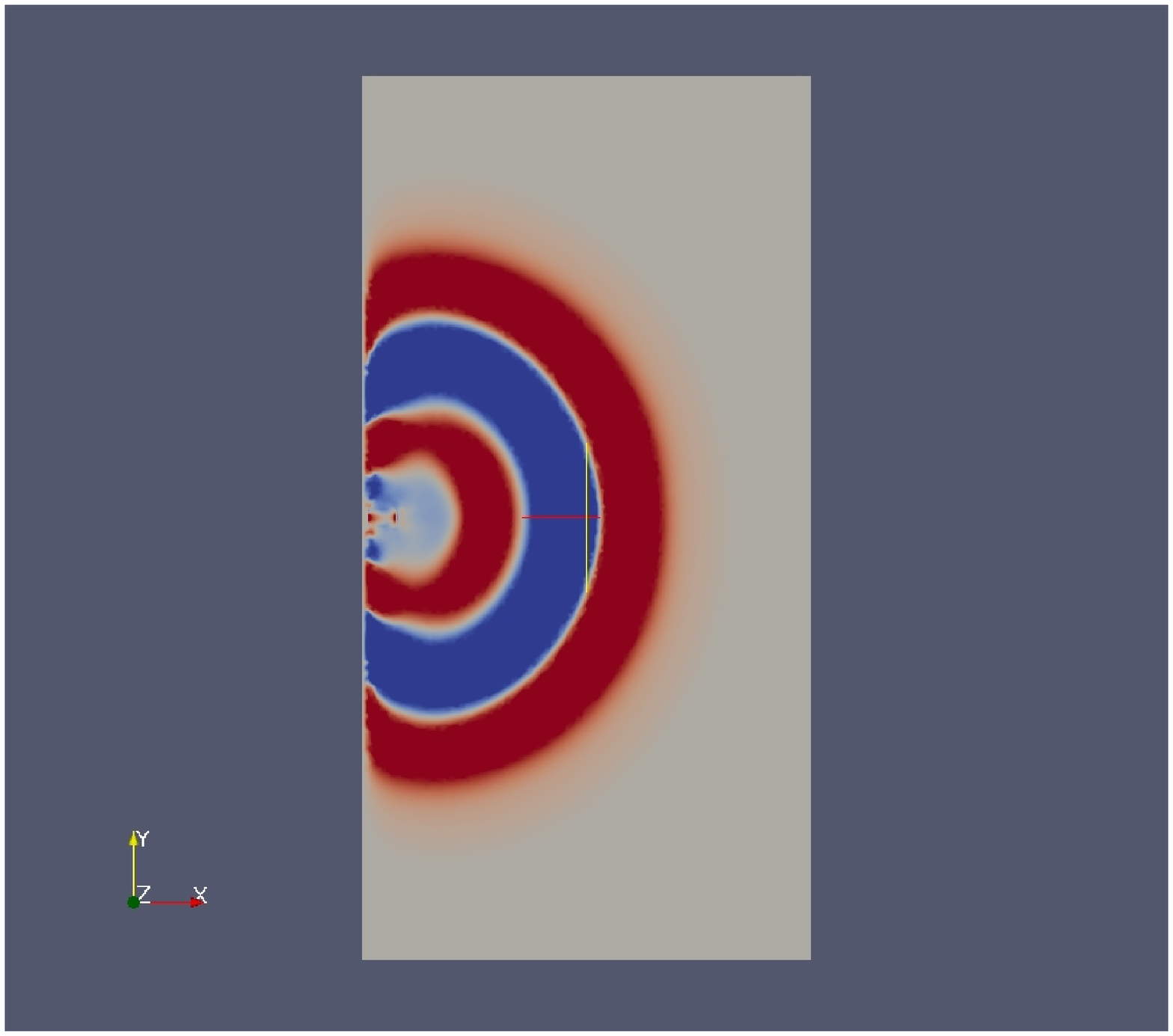}
}
\end{center}
\caption{VOLNA simulations of the Costa Concordia sinking: free surface at $t = 150$ s.}
\label{costa_VOLNA_150s}
\end{figure}

\begin{figure}
\begin{center}
\subfigure[$v = 0.5$ m/s]{
\includegraphics[width=0.31\textwidth]{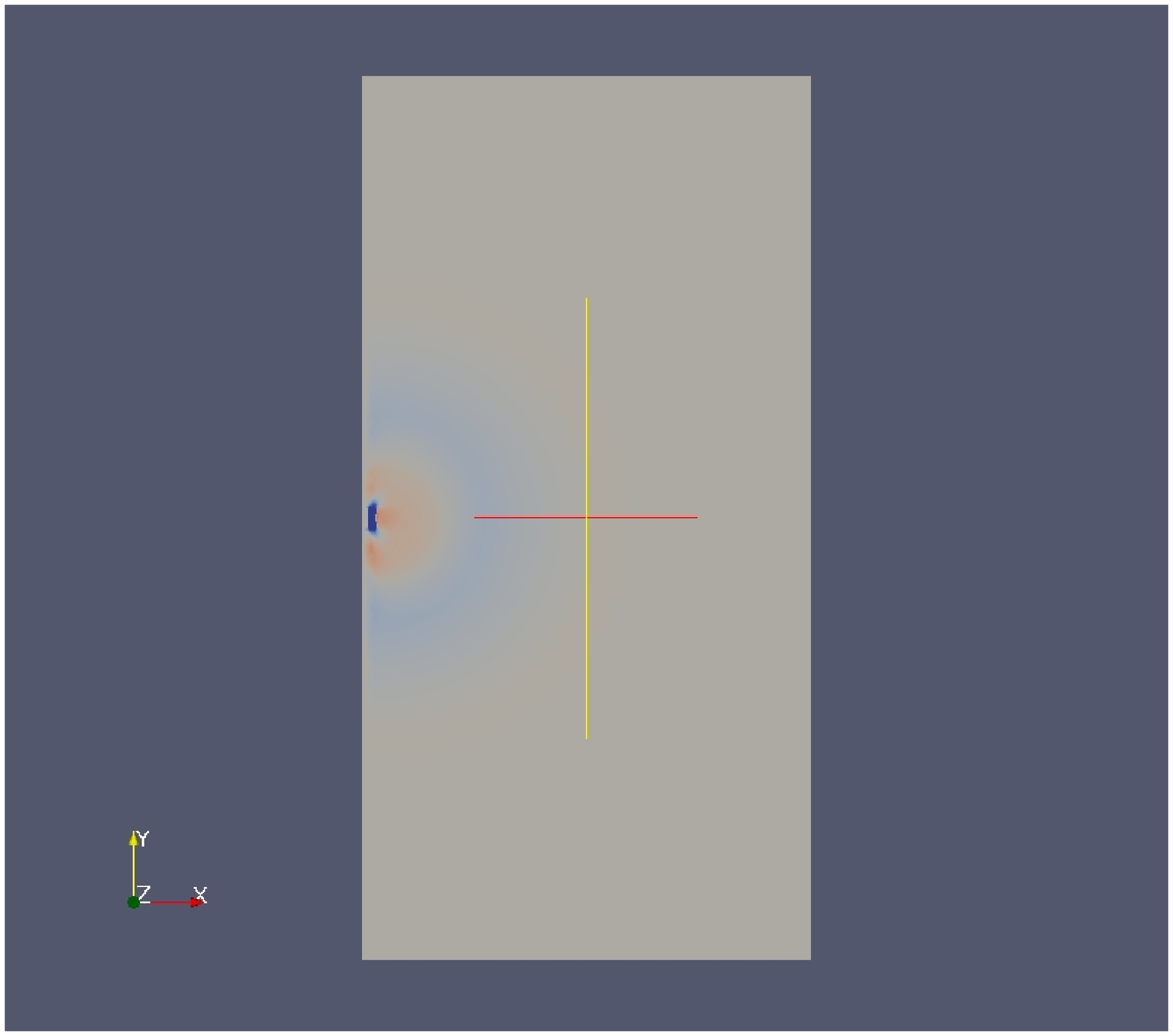}
}
\subfigure[$v = 1$ m/s]{
\includegraphics[width=0.31\textwidth]{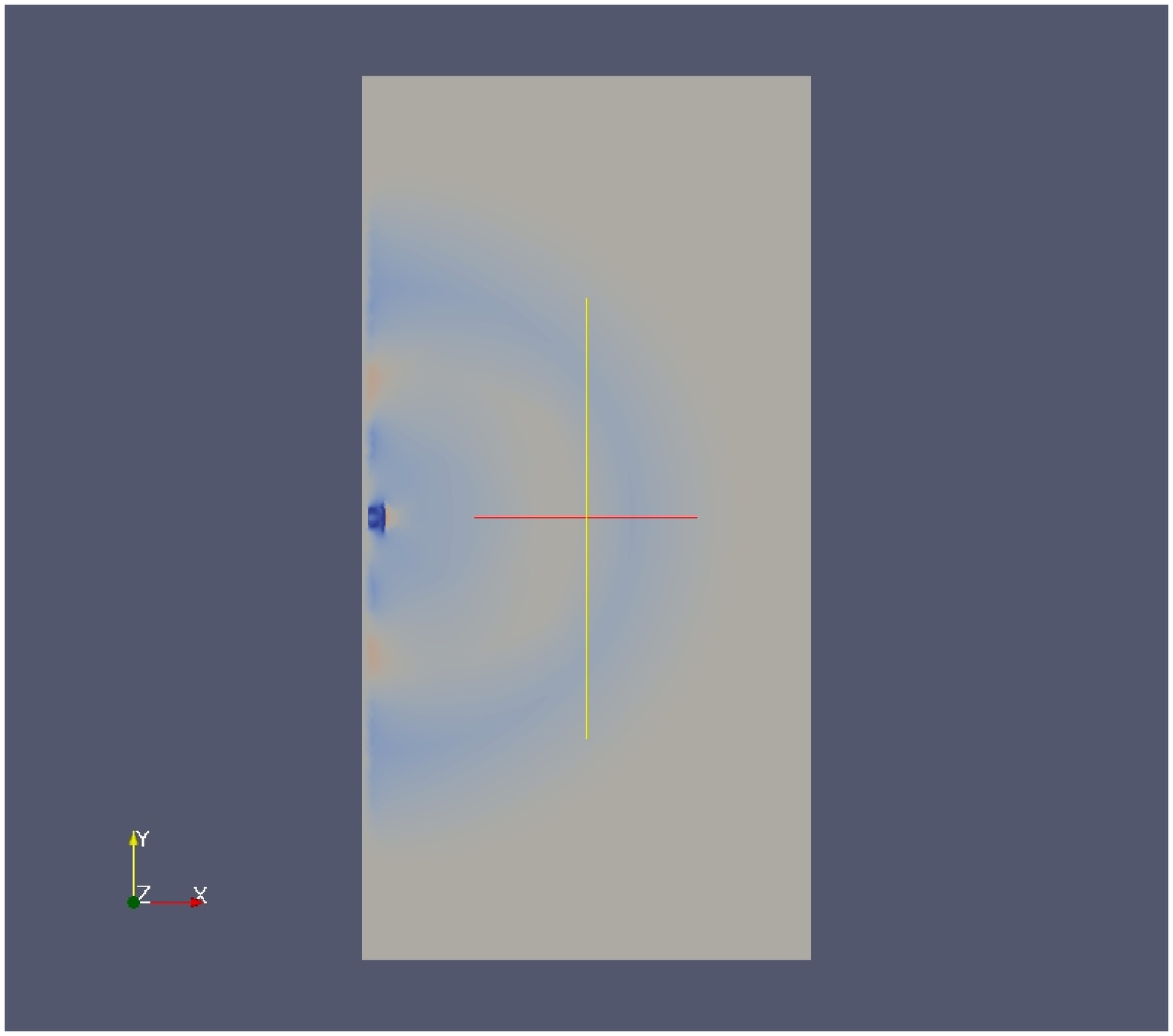}
}
\subfigure[$v = 10$ m/s]{
\includegraphics[width=0.31\textwidth]{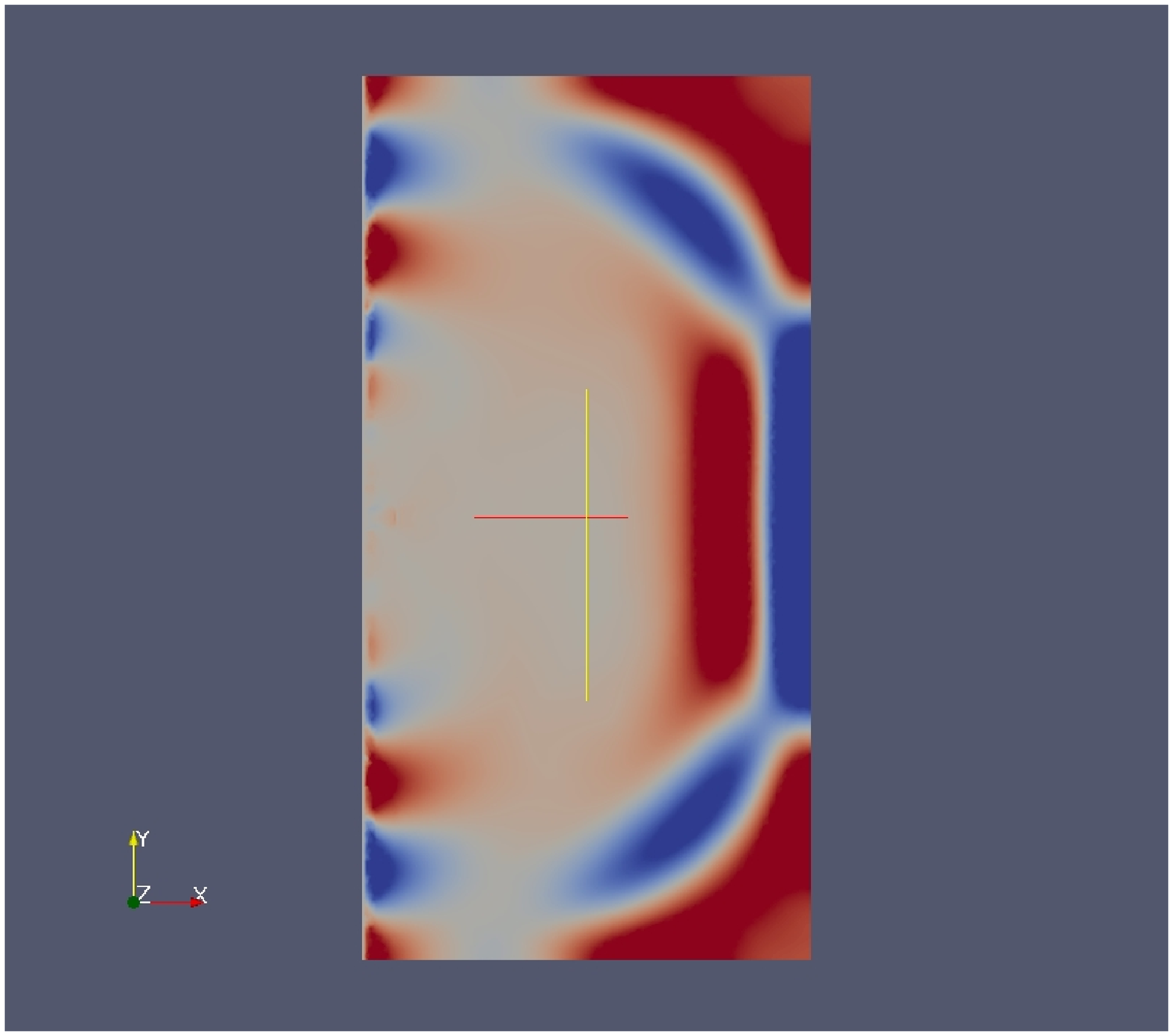}
}
\end{center}
\caption{VOLNA simulations of the Costa Concordia sinking: free surface at $t = 300$ s.}
\label{costa_VOLNA_300s}
\end{figure}

Each simulation was run for $300$ seconds. The free surface plots for each simulation are shown at $t = 75$ s, $t = 150$ s and $t = 300$ s in Figures \ref{costa_VOLNA_75s}, \ref{costa_VOLNA_150s} and \ref{costa_VOLNA_300s} respectively.
 
\begin{figure}
\begin{center}
\includegraphics[width=0.7\textwidth]{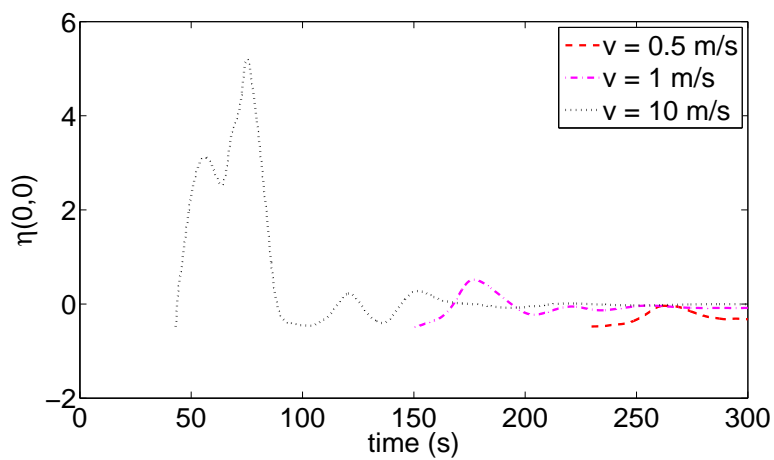}
\caption{Wave gauge at $(x,y) = (0,0)$ for VOLNA simulations of the Costa Concordia sinking from time when the ship has left the shore.}
\label{costa_shore_gauge}
\end{center}
\end{figure}

Wave gauges were recorded at three points for each simulation: at the shore where the ship sinks $(x,y) = (0,0)$, on the shore line far from the slide $(x,y) = (0,400)$ and far offshore from the ship at $(x,y) = (1000,0)$. Figure \ref{costa_shore_gauge} shows the free surface at $(x,y) = (0,0)$ from the time the ship leaves the shore and is completely submerged. The free surface values before this time are not applicable because the water depth is zero. For $v = 10$ m/s the maximum run-up value is $5.2$ m, for $v = 1$ m/s the maximum run-up is $0.52$ m and for $v = 0.5$ m/s there is no run-up for $t<300$ s.

\begin{figure}
\begin{center}
\subfigure[]{
\includegraphics[width=0.47\textwidth]{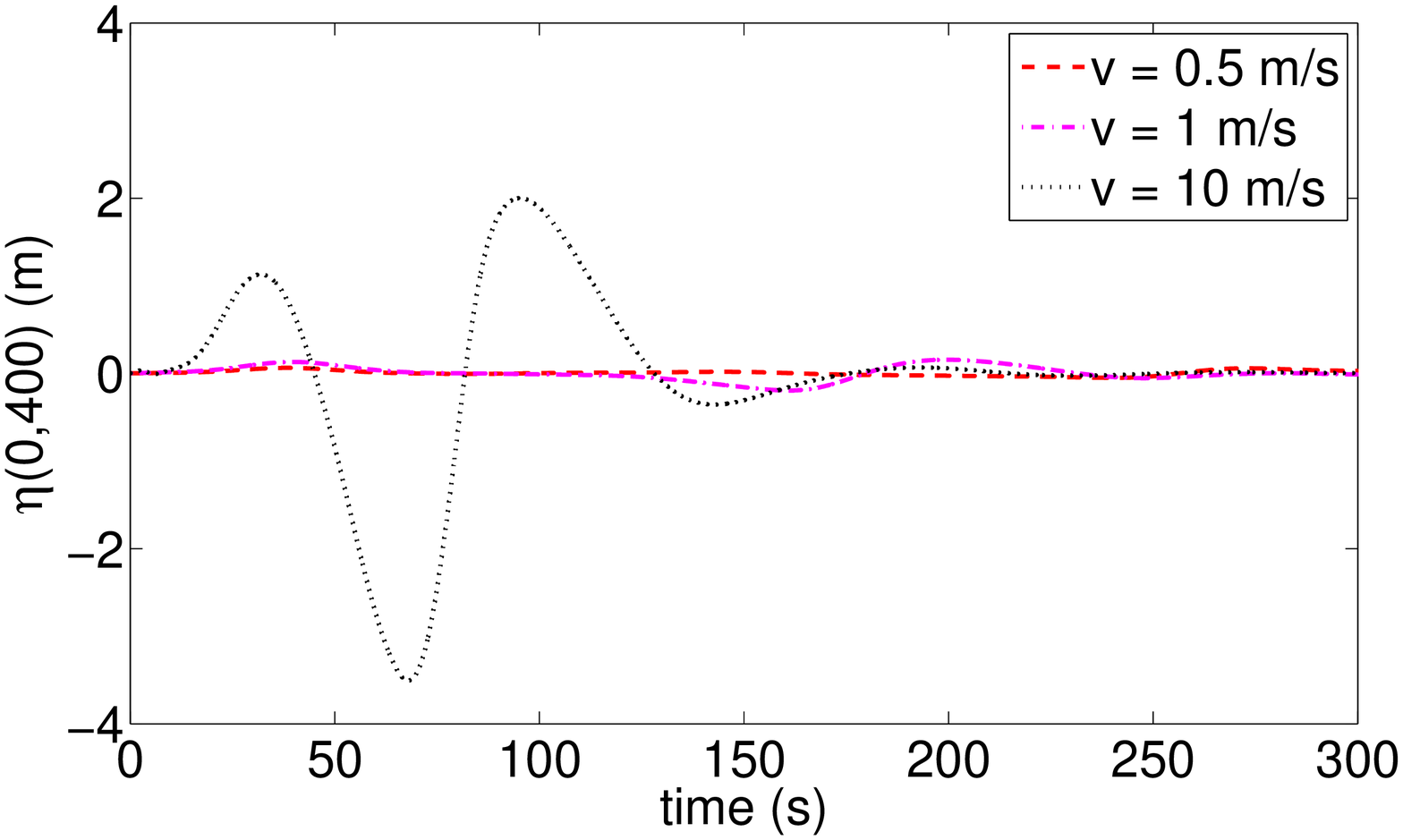} 
}
\subfigure[]{
\includegraphics[width=0.47\textwidth]{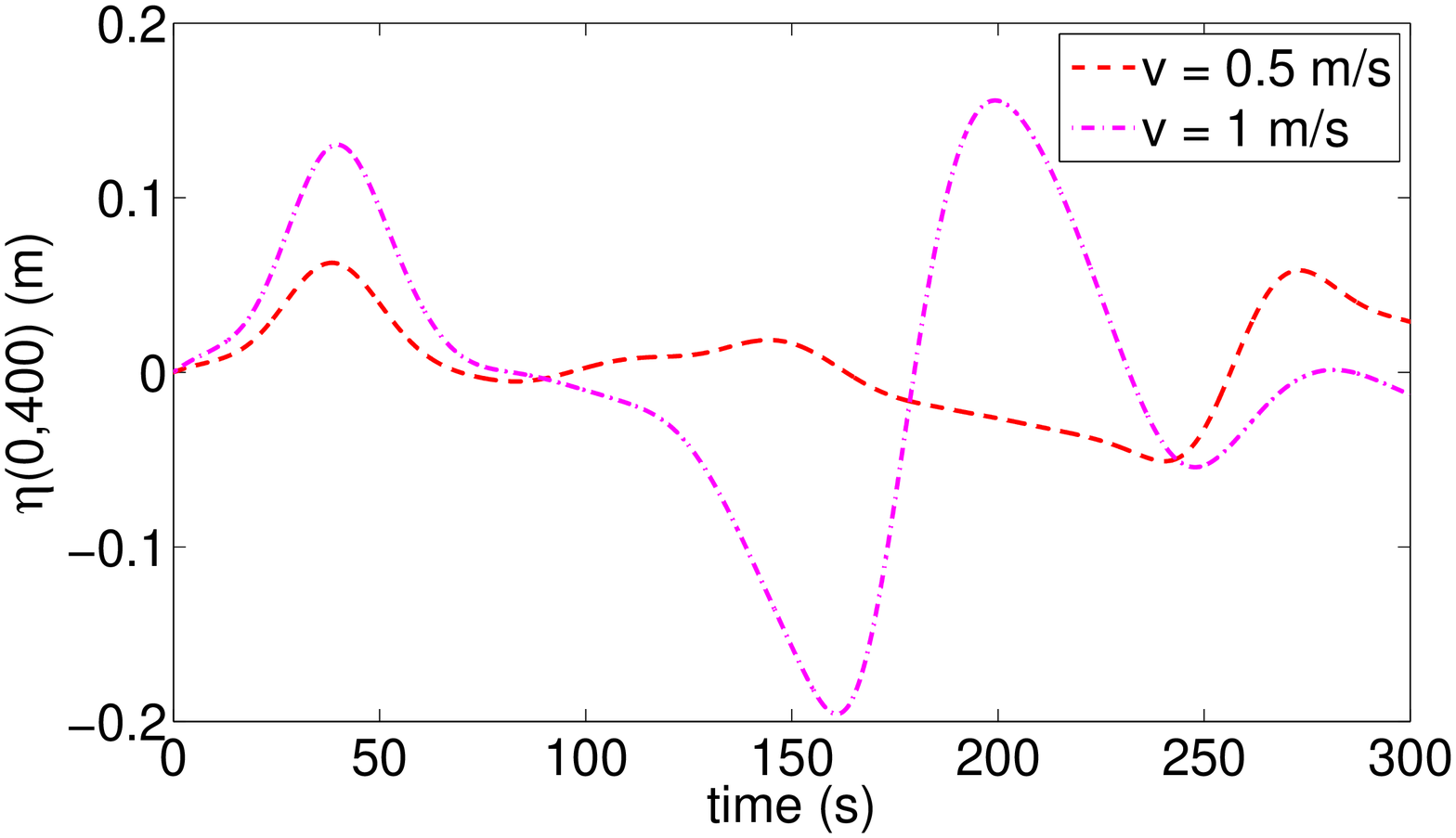} 
}
\end{center}
\caption{Wave gauge at $(x,y) = (0,400)$ for VOLNA simulations of the Costa Concordia sinking: (a) $v = 0.5$ m/s, $v = 1$ m/s and $v = 10$ m/s; (b) close up of $v = 0.5$ m/s and $v = 1$ m/s}.
\label{costa_edge_gauge}
\end{figure}

Figure \ref{costa_edge_gauge} shows the wave gauge on the shore line far from the slide at $(x,y) = (0,400)$. Clearly the free surface amplitude for $v = 10$ m/s is much larger than for $v = 1$ m/s and $v = 0.5$ m/s. The maximum run-up for $v = 10$ m/s is $2$ m, and negligible for the other two velocities.

\begin{figure}
\begin{center}
\subfigure[]{
\includegraphics[width=0.47\textwidth]{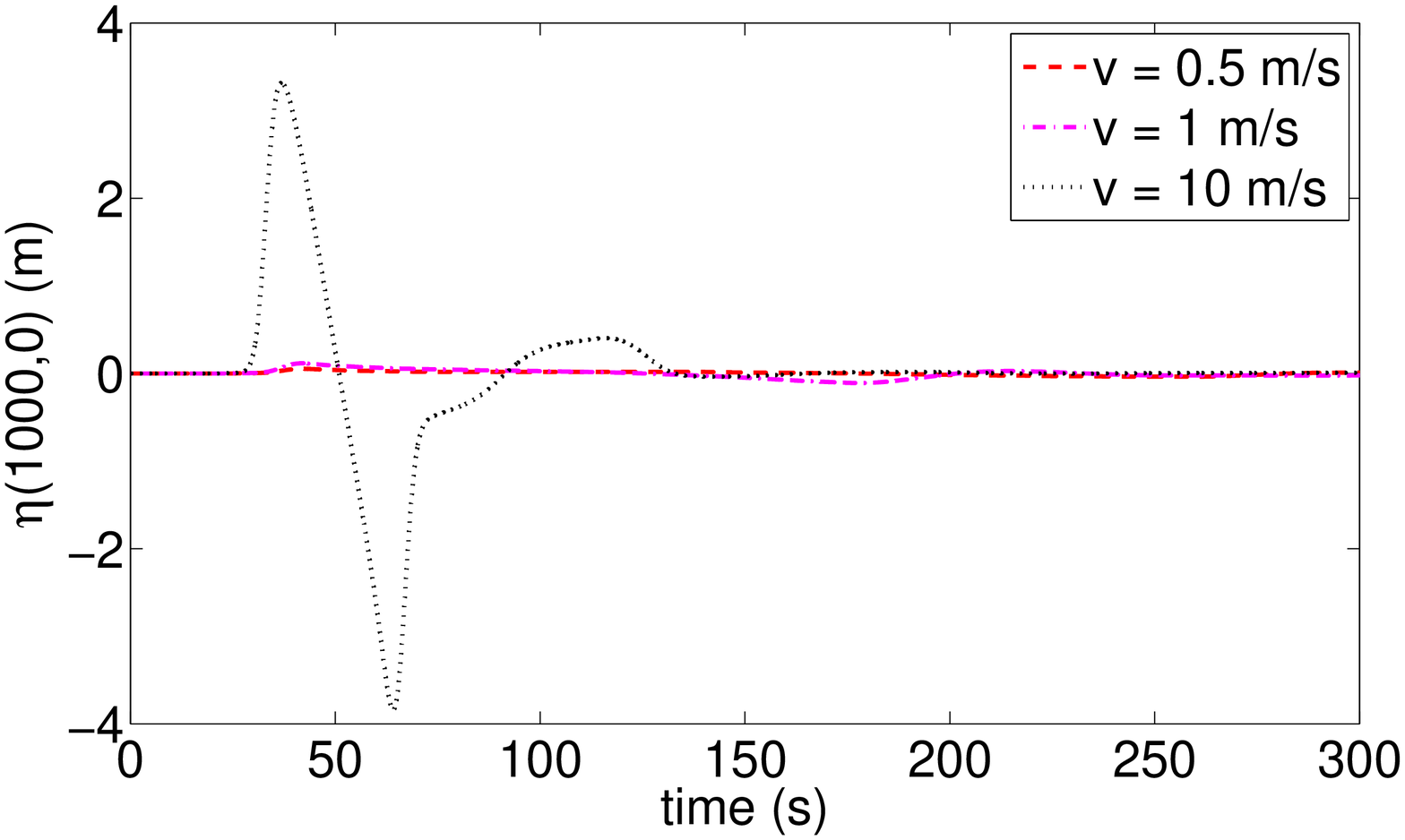} 
}
\subfigure[]{
\includegraphics[width=0.47\textwidth]{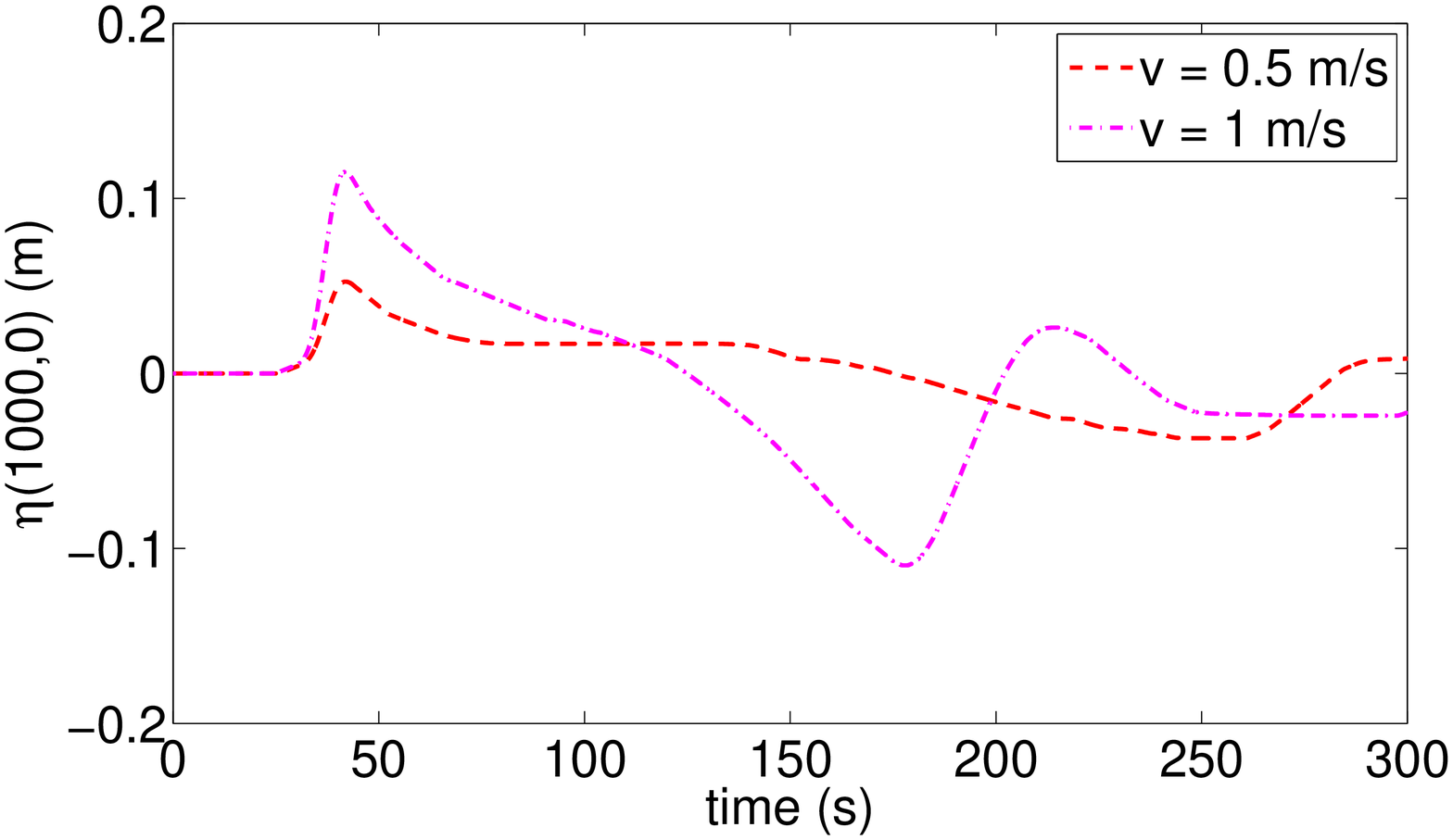} 
}
\end{center}
\caption{Wave gauge at $(x,y) = (1000,0)$ for VOLNA simulations of the Costa Concordia sinking: (a) $v = 0.5$ m/s, $v = 1$ m/s and $v = 10$ m/s; (b) close up of $v = 0.5$ m/s and $v = 1$ m/s}.
\label{costa_far_gauge}
\end{figure}

Figure \ref{costa_far_gauge} shows the wave gauge far offshore from the ship at $(x,y) = (1000,0)$. The maximum run-up values for $v = 10, 1$ and $0.5$ m/s are $3.32, 0.12$ and $0.05$ m respectively. Once again these results demonstrate the effect of the sinking velocity. If the ship were to sink slowly then a wave of only a few centimeters would be generated, however if it were to slide quickly down the steep slope it lies on, waves of a few meters could threaten not only the immediate coast, but the Italian mainland and the coastline on the island north and south of the ship. In particular, a wave entering the nearby port south of the Costa Concordia could be reflected and amplified.

Since the sinking velocity is an important factor in determining the size of the waves, it is important to know the average density of the ship. Reports on the weight of the Costa Concordia vary. She has a dead weight of $10,000$ t but some reports suggest that she has an actual weight of $45,000$ t, not including luggage or water inside. If the ship is half filled with water then the mass of the water is $2.88$ x $10^8$ kg and adding this to $45,000$ t would give the ship a density of $579$ kg/m$^3$. This would give $\rho_s/\rho_w < 1$. However, if the ship was filled with water then $\rho_s = 1078$ kg/m$^3$ and  $\rho_s/\rho_w \approx 1$. Also air pockets within the ship would effect the sinking kinematics.

In conclusion, if the Costa Concordia were to suddenly and quickly slip down the steep slope she lies on then large waves could be generated and pose a danger to the immediate, far field and offshore coastlines. However, given that this is a ship and not a solid mass it is likely that the sinking kinematics would not be straight forward. It is difficult to determine the density of the ship and how air pockets may effect the way in which she would sink.

\section{Tsunami inundation}\label{sec:inun}

When it comes to tsunami inundation, one could simply say: rely on the wisdom of ancient people. In Japan, for example, hundreds of stone tablets warn citizen about the dangers of a tsunami. For example, one tablet in Aneyoshi, a small coastal town, reads: ''High dwellings are the peace and harmony of our descendants. Remember the calamity of the great tsunami. Do not build any homes below this point.'' As the ancient warnings in stone attest, tsunamis are not new to this vulnerable part of Japan. In 1896, during the Meiji Period, a tsunami killed at least 22,000 people on the same Sanriku Coast - a death toll chillingly close to the recent 2011 disaster which left an estimated 23,000 dead or missing.

\subsection{Run-up amplification}

\begin{figure}
\begin{center}
\includegraphics[width=0.9\textwidth]{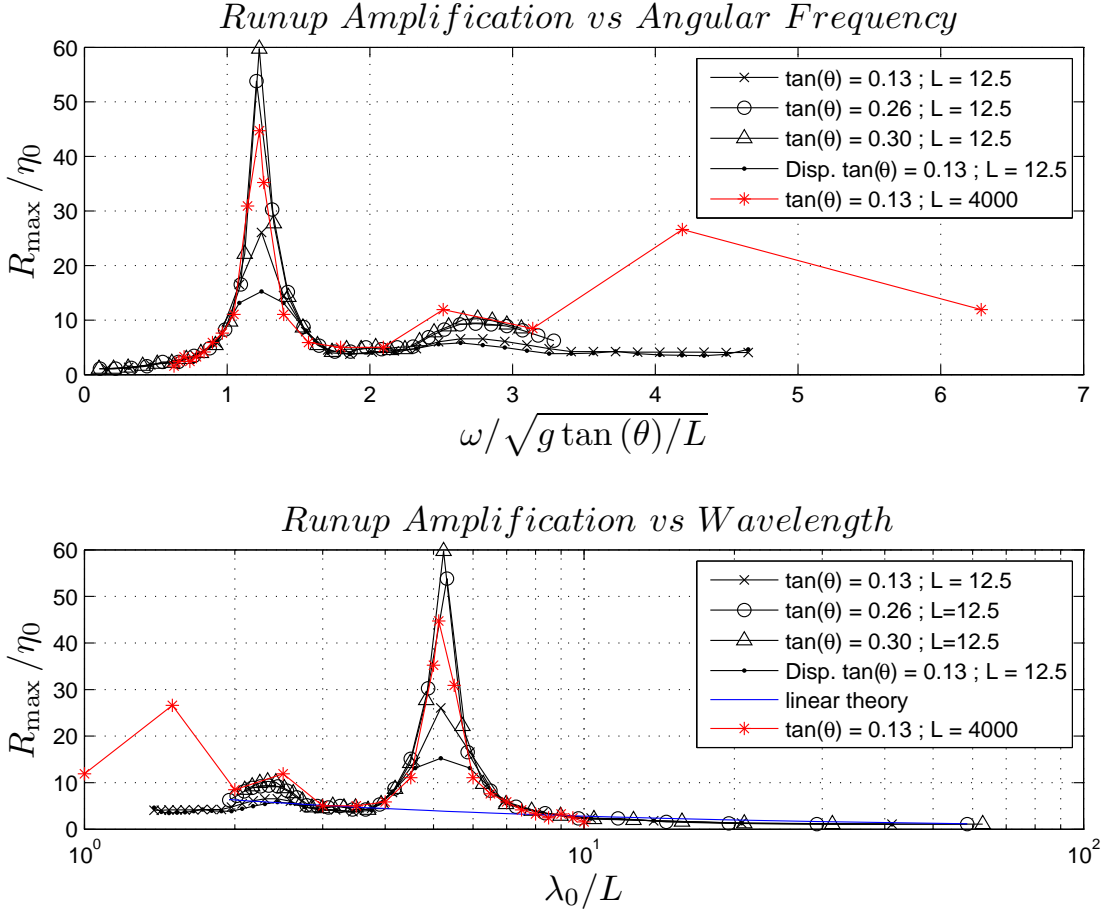}
\caption{Maximum run-up amplification ratio as a function of non-dimensional angular frequency (top) and non-dimensional wavelength (bottom) for two beach lengths, namely $L=12.5$ m and $4000$ m (from \cite{Stefanakis2011}).}
\label{fig:sin_ampli}
\end{center}
\end{figure}

Until recently the analysis of long wave run-up on a plane beach has been focused on finding its maximum value, failing to capture the existence of resonant regimes. \textsc{Stefanakis} \emph{et al}. \cite{Stefanakis2011} performed one-dimensional numerical simulations in the framework of the NSWE to investigate the Boundary Value Problem (BVP) for plane and non-trivial beaches. Monochromatic waves, as well as virtual wave-gage recordings from real tsunami simulations, were used as forcing conditions to the BVP. Resonant phenomena between the incident wavelength and the beach slope were found to occur, which result in enhanced run-up of non-leading waves (these resonances are different from edge-wave resonances and from bay resonances, which are two-dimensional phenomena). Run-up amplification occurs for both leading elevation and depression waves. Figure \ref{fig:sin_ampli} shows the maximum run-up amplification ratio as a function of non-dimensional frequency and non-dimensional wavelength for two beach lengths, namely $L = 12.5$ m and $4000$ m. Here $\eta_0$ is the incoming wave amplitude, $\omega$ the frequency, $g$ the acceleration due to gravity, $\theta$ the beach slope, $\lambda_0$ the incoming wavelength. The classical formula based on linear theory for a constant slope reads (see for example \cite{Madsen2008})
\begin{equation}
\frac{R}{\eta_0} = 2\sqrt{\pi} \left( \frac{L \omega^2}{g \tan\theta} \right)^{1/4}.
\end{equation}
Even though these results are in an ideal setting (plane beach, incoming sinusoidal wave) they have been shown to persist in more realistic situations (arbitrary bathymetry, complex wave signal). The effects of dispersion have been studied as well: dispersion reduces slightly the amplification ratio. An experimental confirmation of the theoretical results of \textsc{Stefanakis} \emph{et al}. \cite{Stefanakis2011} was recently published by \textsc{Ezersky} \emph{et al}. \cite{Ezersky2012}.

\subsection{Whirlpool like effect in tsunamis}

\begin{figure}
\begin{center}
 \includegraphics[width=0.7\textwidth]{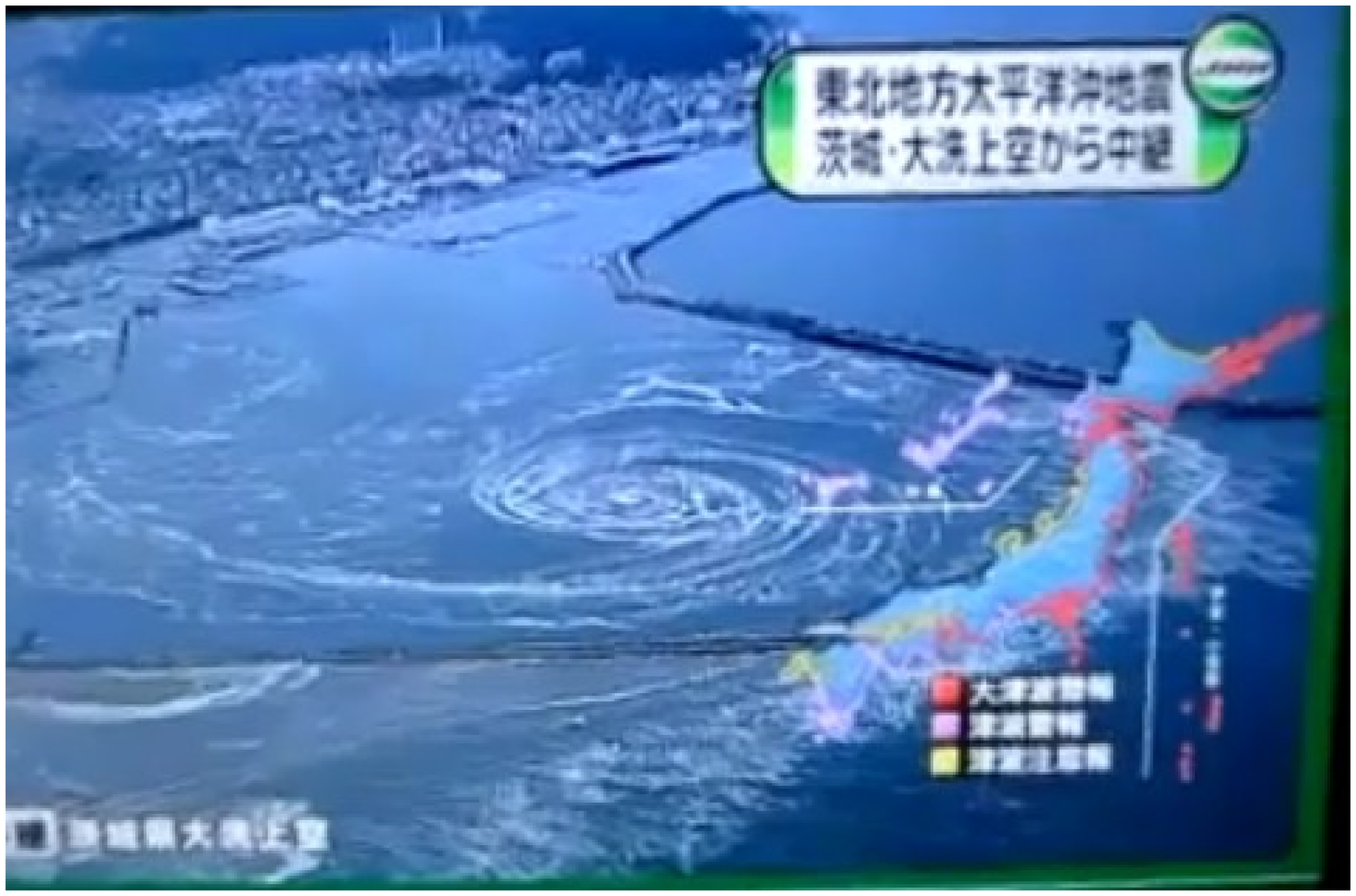}
\caption{Image of Oarai Harbour on a Japanese news channel when the tsunami hit. Source: \url{http://www.youtube.com/watch?v=Qgp1jq6Og4k} (screenshot 10/08/11)}
\label{Oarai_News_picture}
\end{center}
\end{figure}

We conclude this paper by investigating an intriguing phenomenon observed during the 11th March 2011 tsunami in Japan. The behaviour of the tsunami in Oarai port was picked up by reporters as it seemed to create a whirlpool like effect within the bounds of the harbour walls (see Figure \ref{Oarai_News_picture}).

\begin{figure}
\begin{center}
\subfigure[]{
\includegraphics[width=0.47\textwidth]{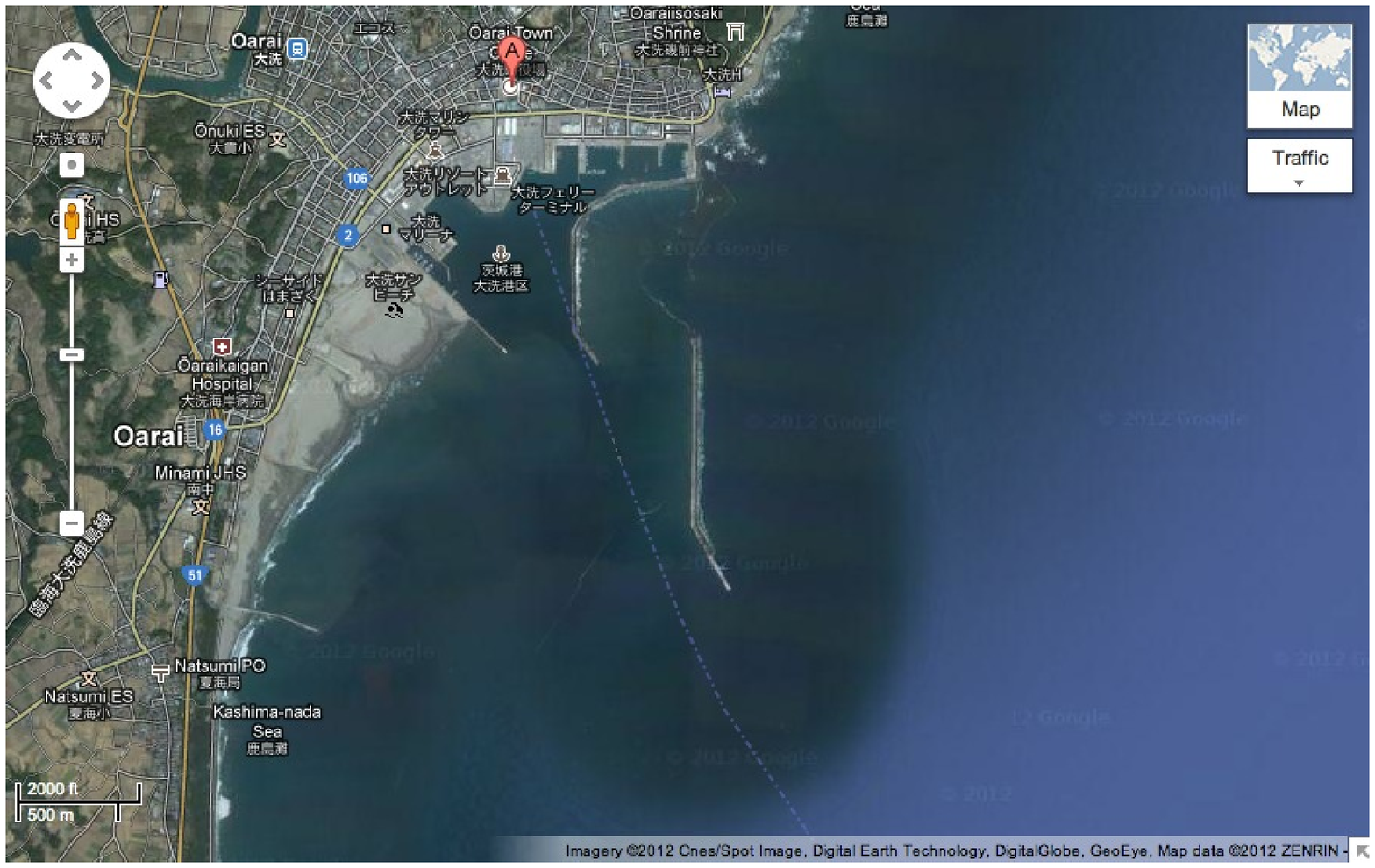}
}
\subfigure[]{
\includegraphics[width=0.47\textwidth]{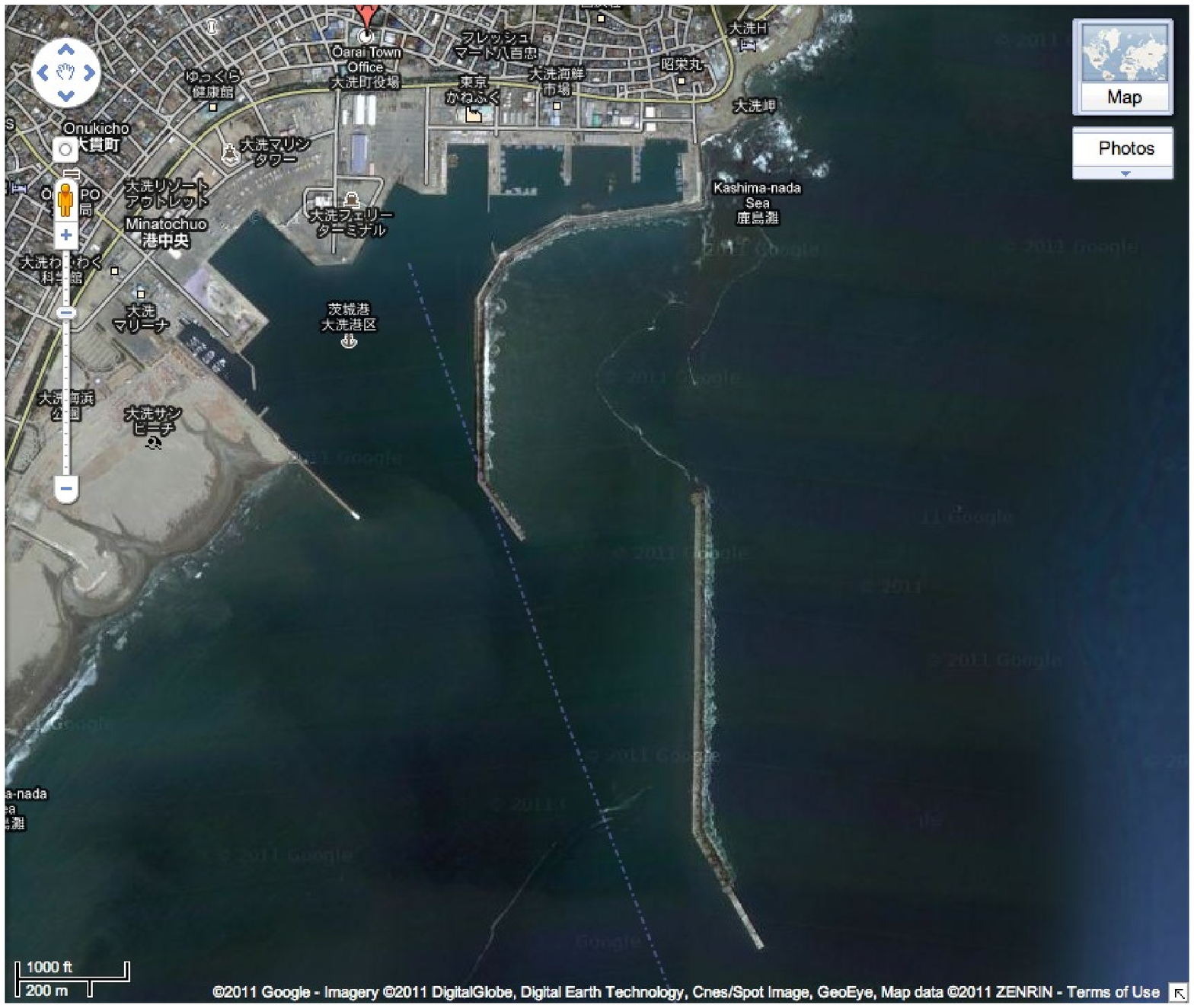}
}
\end{center}
\caption{Satellite images of Oarai harbour from Google Maps (a) far (b) close.}
\label{Oarai_sat_view}
\end{figure}

\begin{figure}
\begin{center}
 \includegraphics[width=0.7\textwidth]{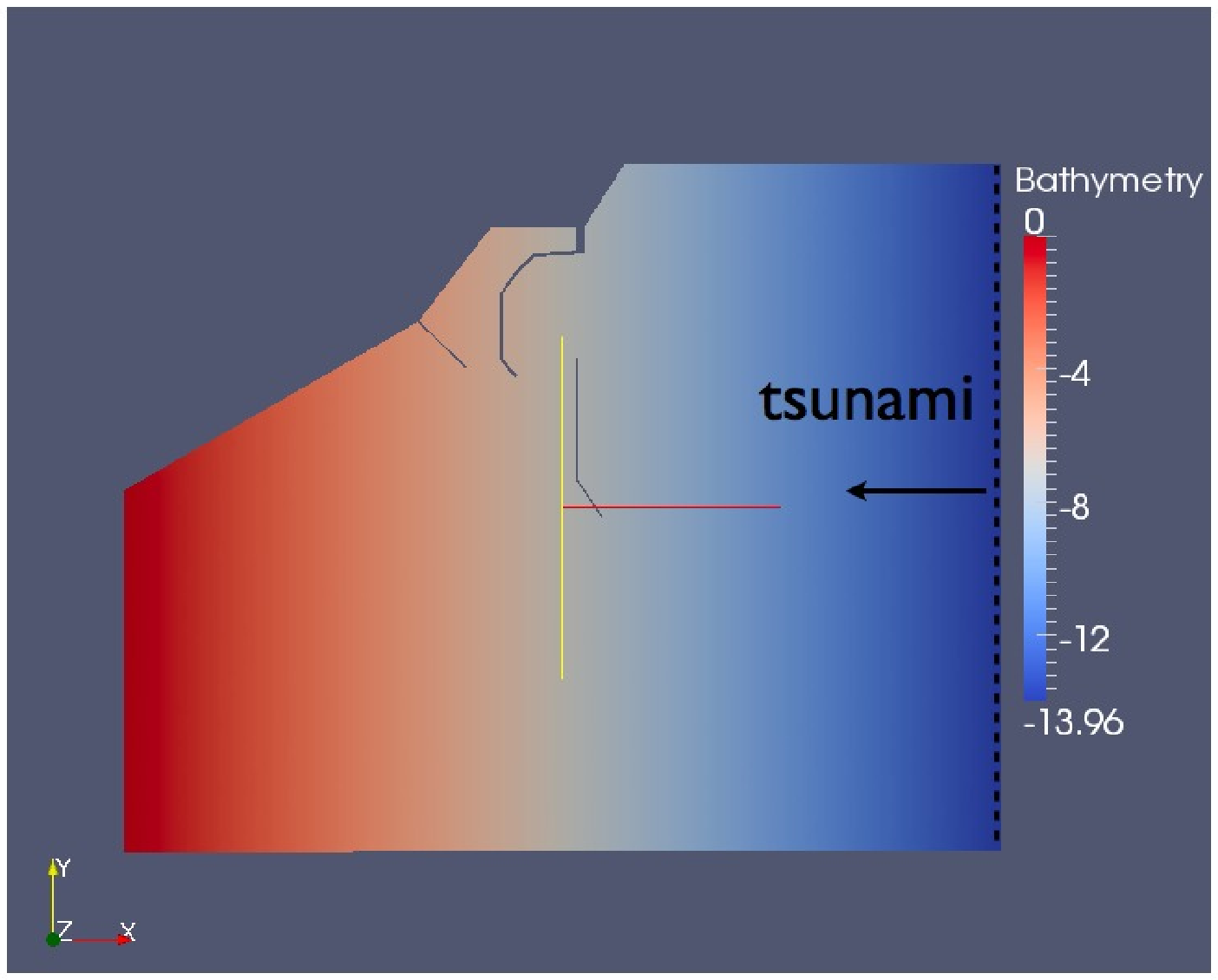}
\caption{Simplified bathymetry used for Oarai harbour simulation, the tsunami is generated at the right boundary and travels from right to left, all other boundaries are wall boundaries.}
\label{Oarai_Bathymetry}
\end{center}
\end{figure}

In an attempt to understand this phenomenon better, the VOLNA code was run on a simplified geometry of the harbour, inputting a tsunami from the right hand side of the domain. The harbour was given a depth of $5$ m at the shore with a gently sloping seafloor with slope $1/300$. These values are based on rough values from Google Earth, as the depth is $5$ m just outside the harbour and it drops off to $15$ m over approximately $3$ km. See Figures \ref{Oarai_sat_view} and \ref{Oarai_Bathymetry} for a comparison of the real and simplified geometry. The tsunami was modelled by a sine wave with amplitude $1$ m and period $10.5$ minutes
\[
\eta = -\sin(0.01 t), \qquad \mathrm{at} \quad x = 7000 \, \mathrm{m}.
\]

\begin{figure}
\begin{center}
 \includegraphics[width=0.7\textwidth]{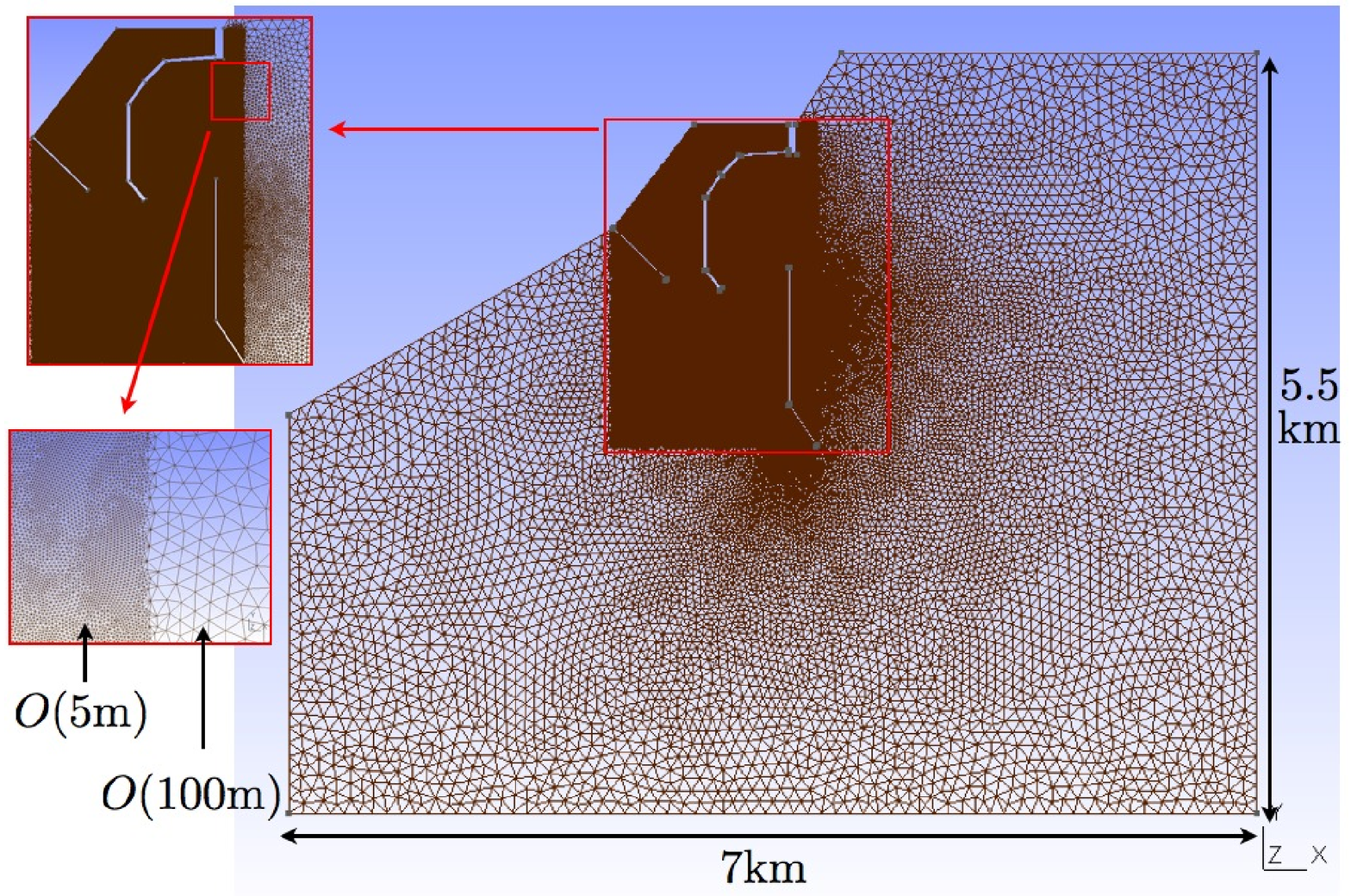}
\caption{Unstructured triangular mesh used for Oarai harbour simulation. There are approximately 320 x 10$^3$ elements of order 100 m, refined to 5 m within the harbour area.}
\label{Oarai_mesh}
\end{center}
\end{figure}

An unstructured triangular mesh was generated over an approximate area of $7$ x $5.5$ km$^2$. The mesh contained approximately 320 x 10$^3$ elements of order $100$ m outside the harbour and refined to $5$ m within the harbour area as shown in Figure \ref{Oarai_mesh}.

\begin{figure}
\begin{center}
\subfigure[$t = 7$ min]{
\includegraphics[width=0.31\textwidth]{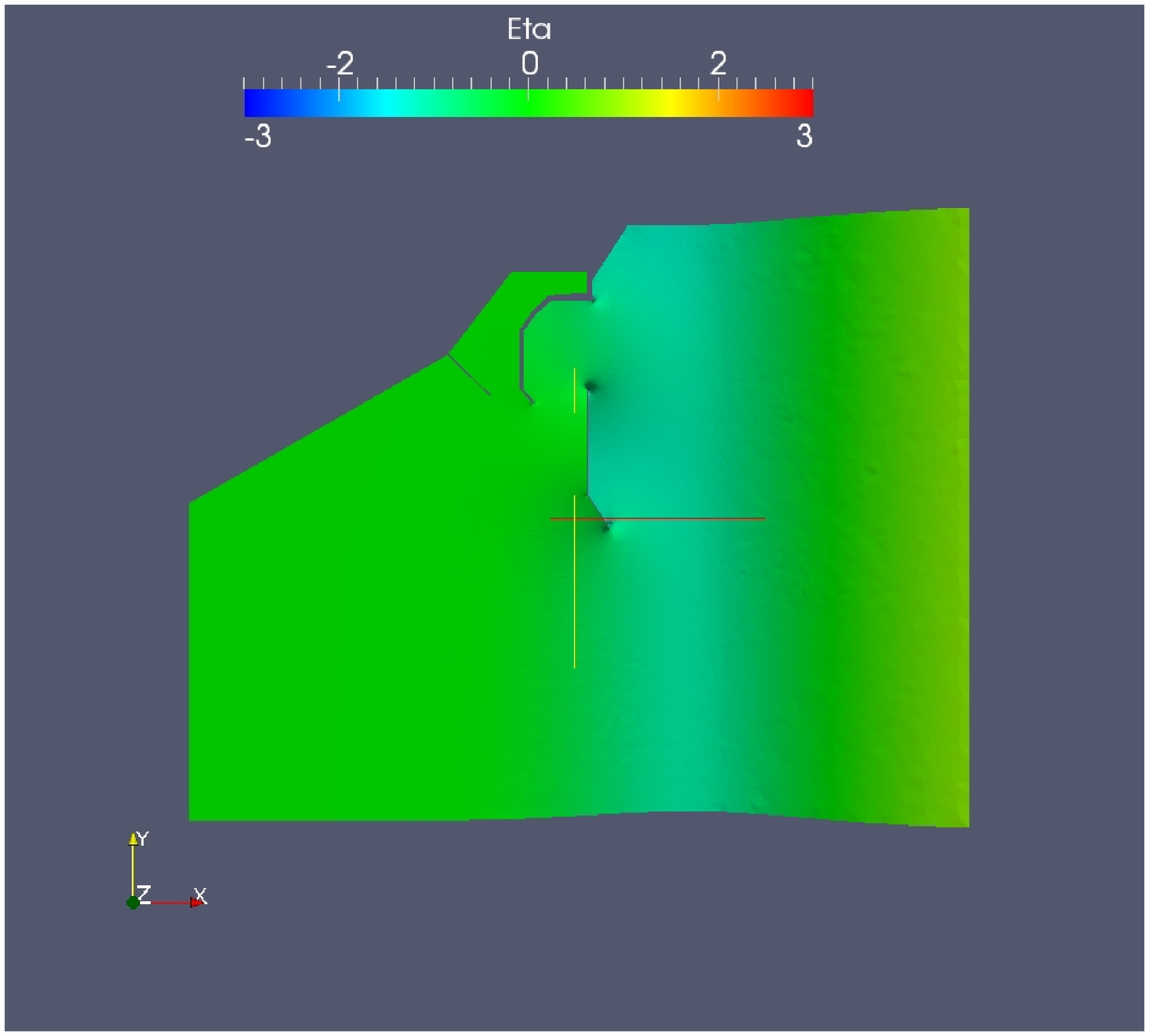}
}
\subfigure[$t = 13$ min]{
\includegraphics[width=0.31\textwidth]{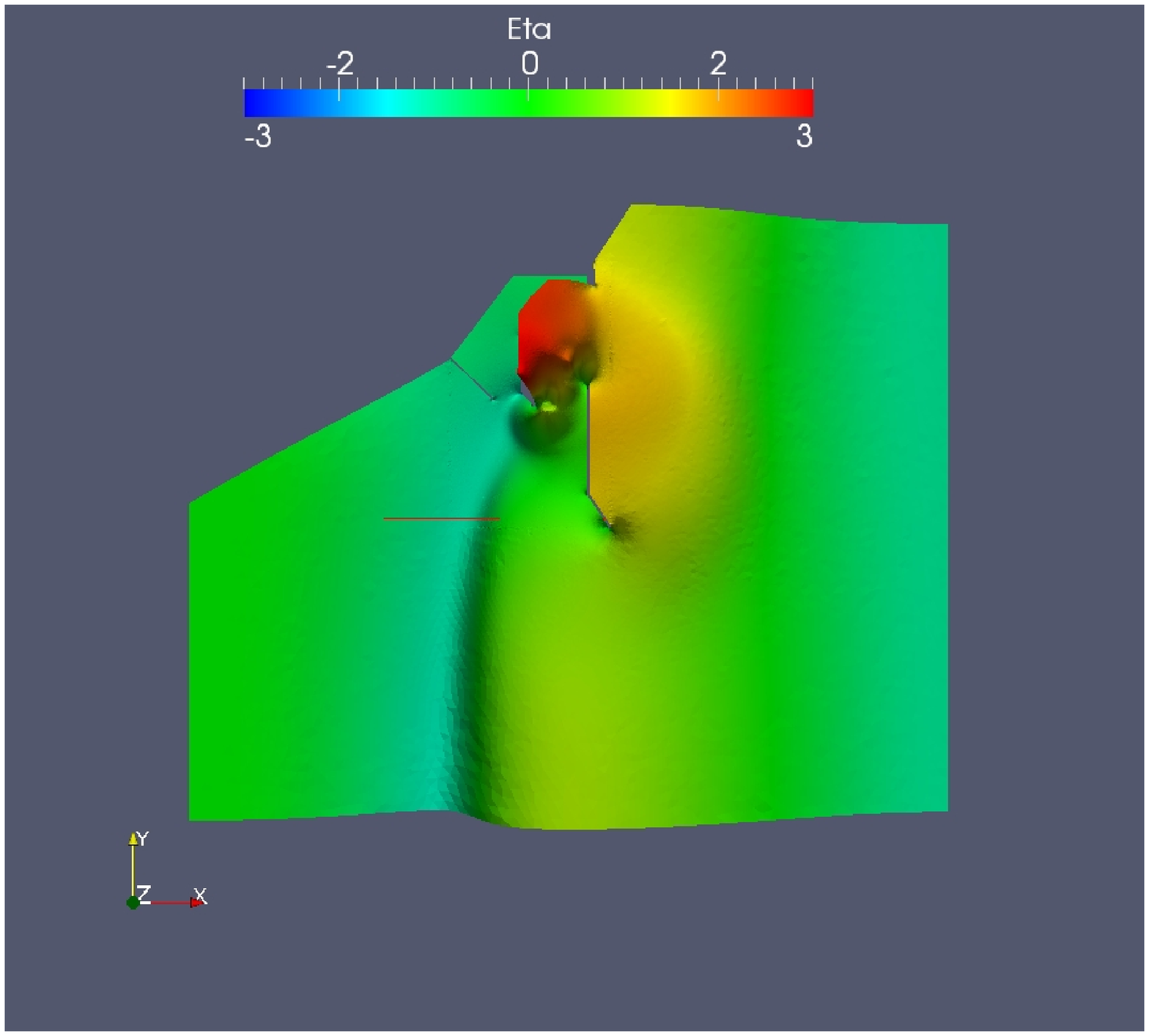}
}
\subfigure[$t = 14$ min]{
\includegraphics[width=0.31\textwidth]{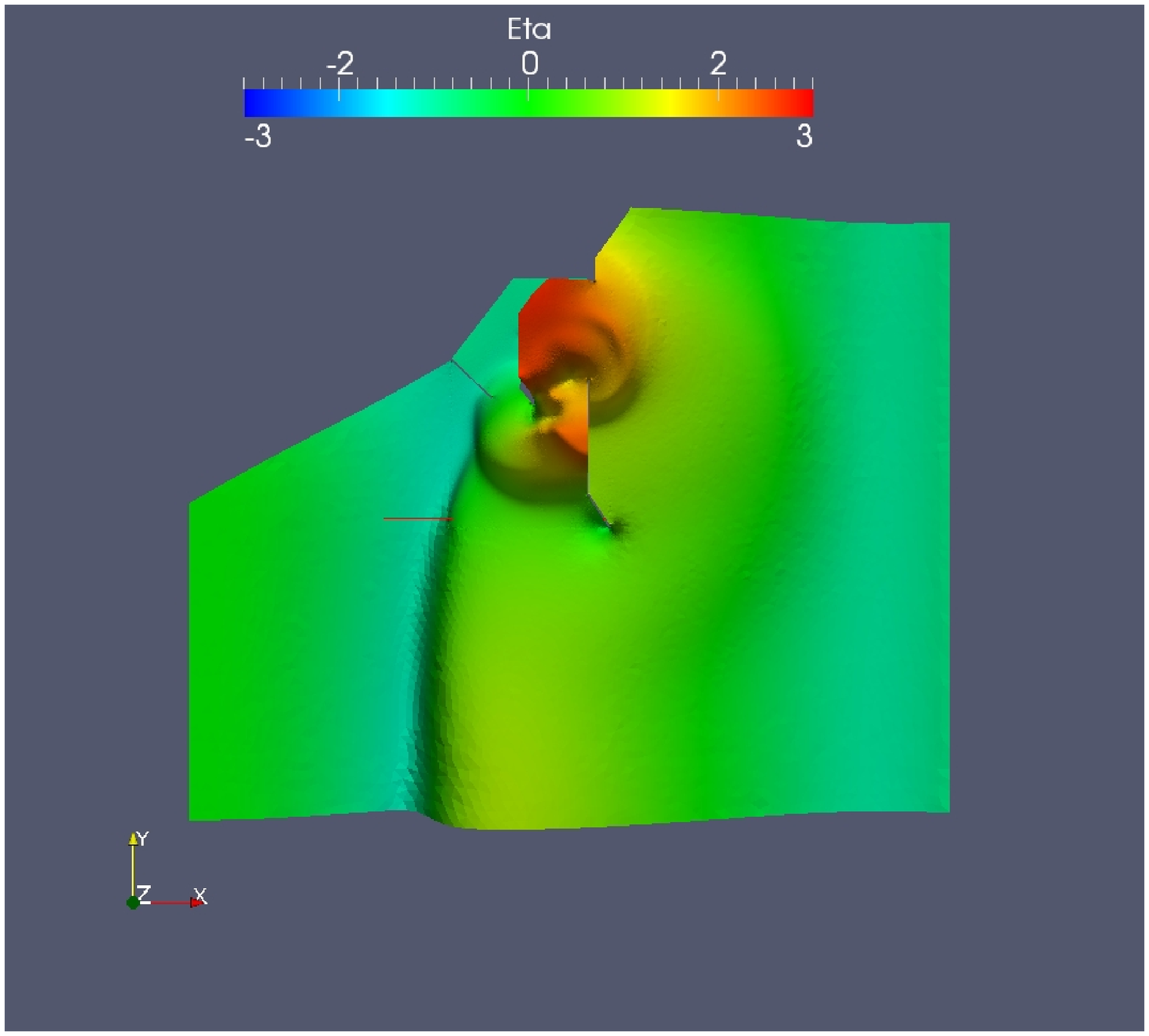}
}
\subfigure[$t = 15$ min]{
\includegraphics[width=0.31\textwidth]{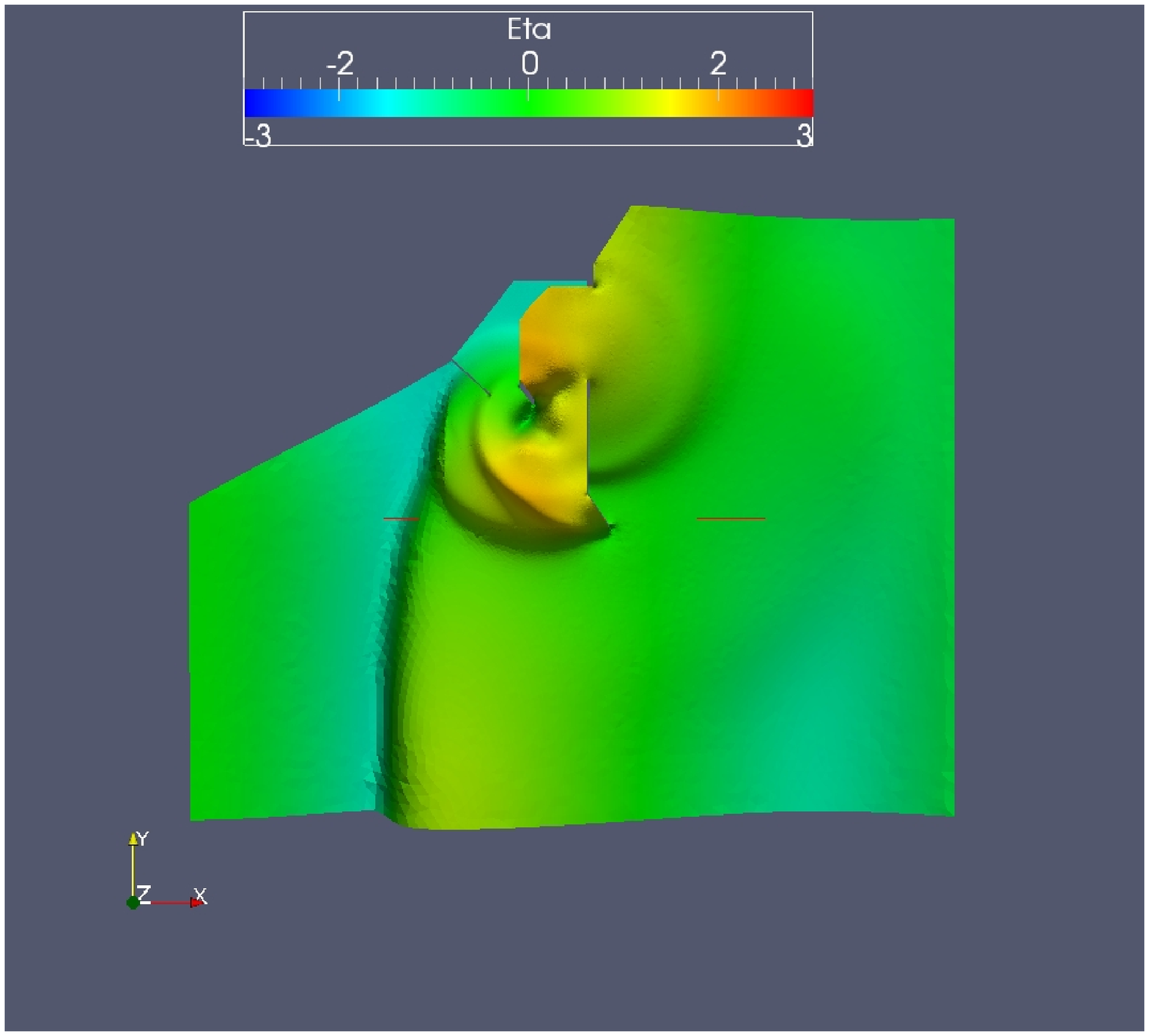}
}
\subfigure[$t = 16$ min]{
\includegraphics[width=0.31\textwidth]{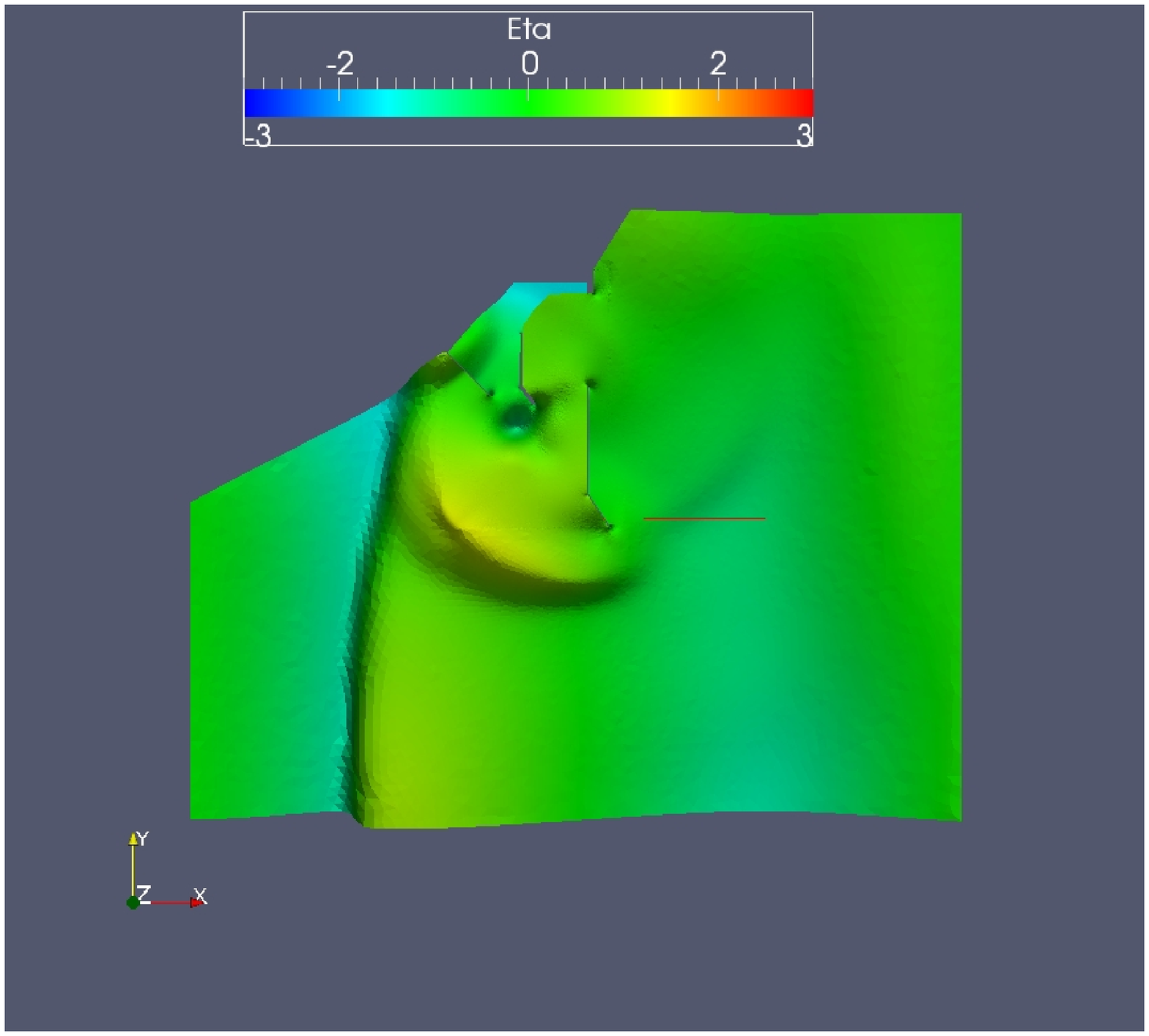}
}
\subfigure[$t = 18$ min]{
\includegraphics[width=0.31\textwidth]{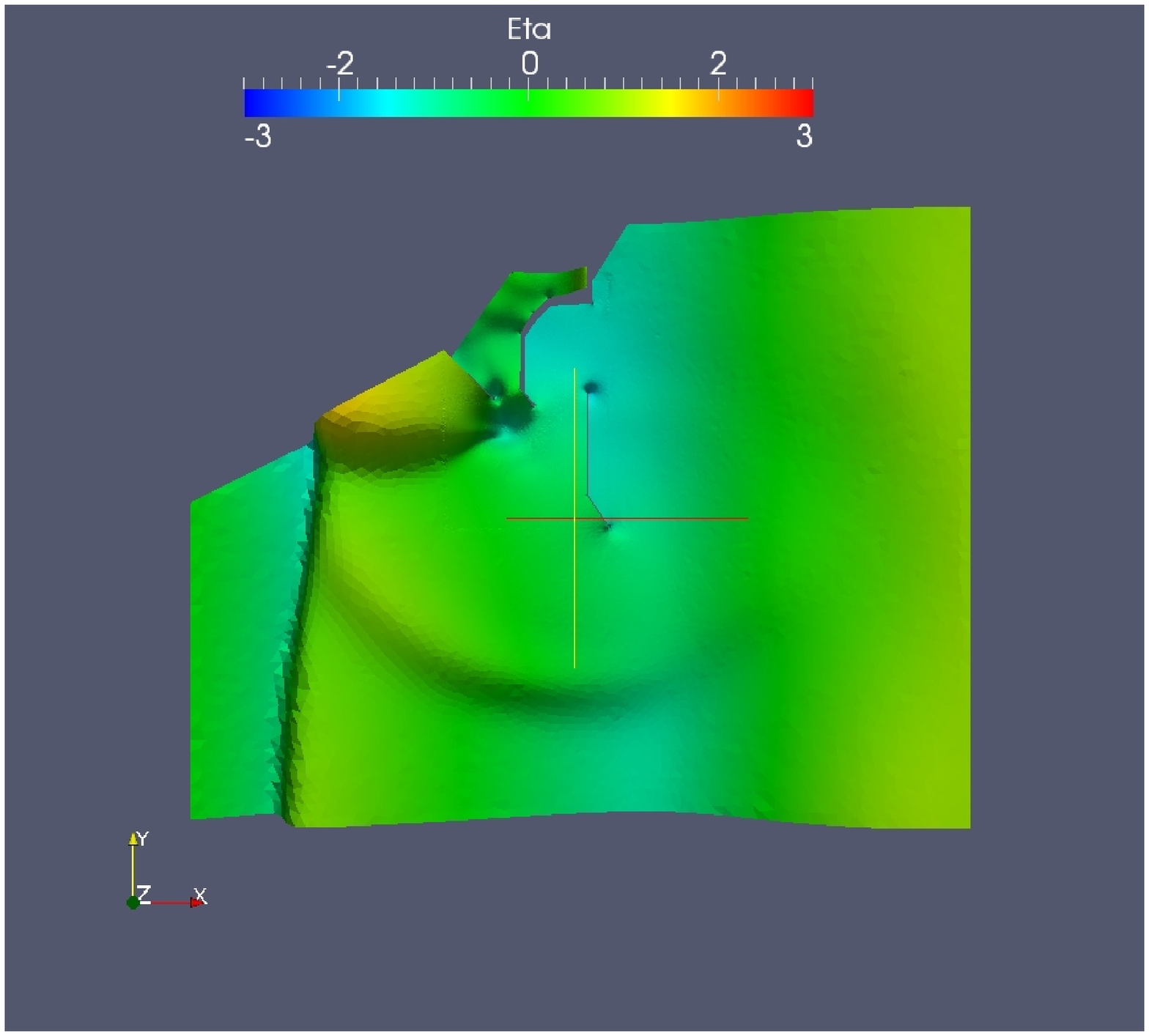}
}
\end{center}
\caption{Snapshots of the free surface between $7$ and $18$ minutes.}
\label{Oarai_results}
\end{figure}

\begin{figure}
\begin{center}
\subfigure[$t = 13$ min]{
\includegraphics[width=0.31\textwidth]{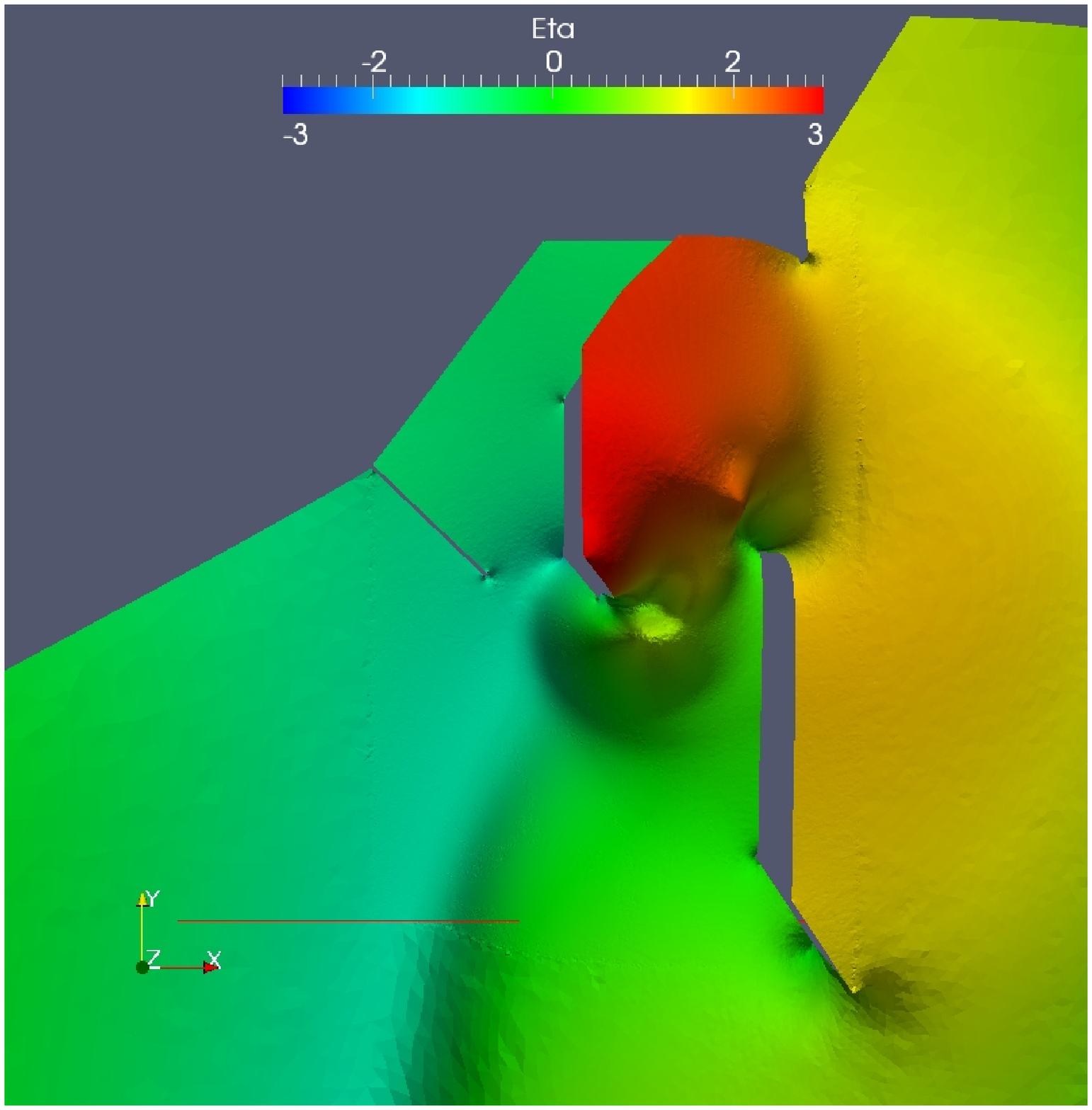} 
}
\subfigure[$t = 13$ min $20$ sec]{
\includegraphics[width=0.31\textwidth]{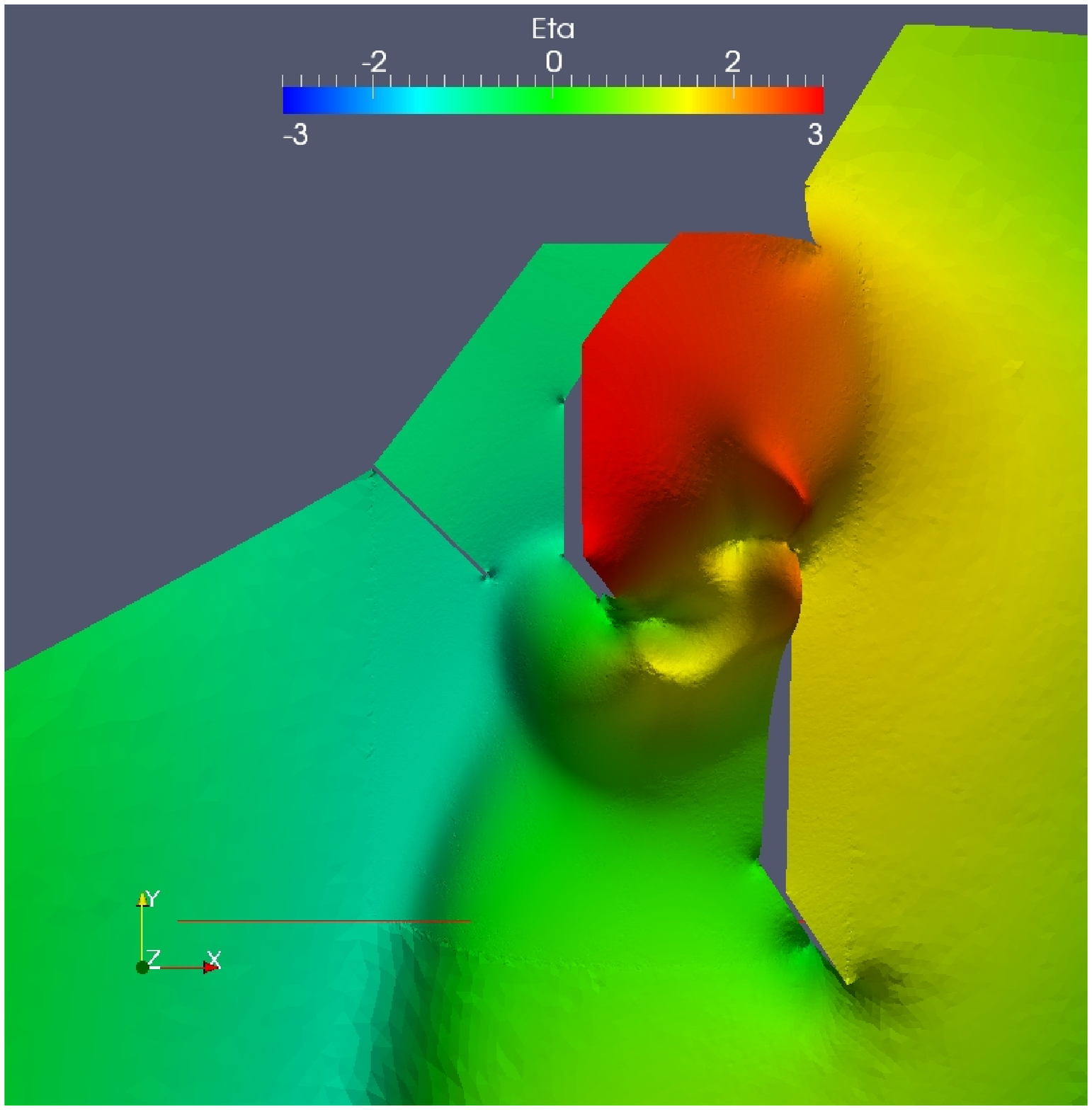} 
}
\subfigure[$t = 13$ min $40$ sec]{
\includegraphics[width=0.31\textwidth]{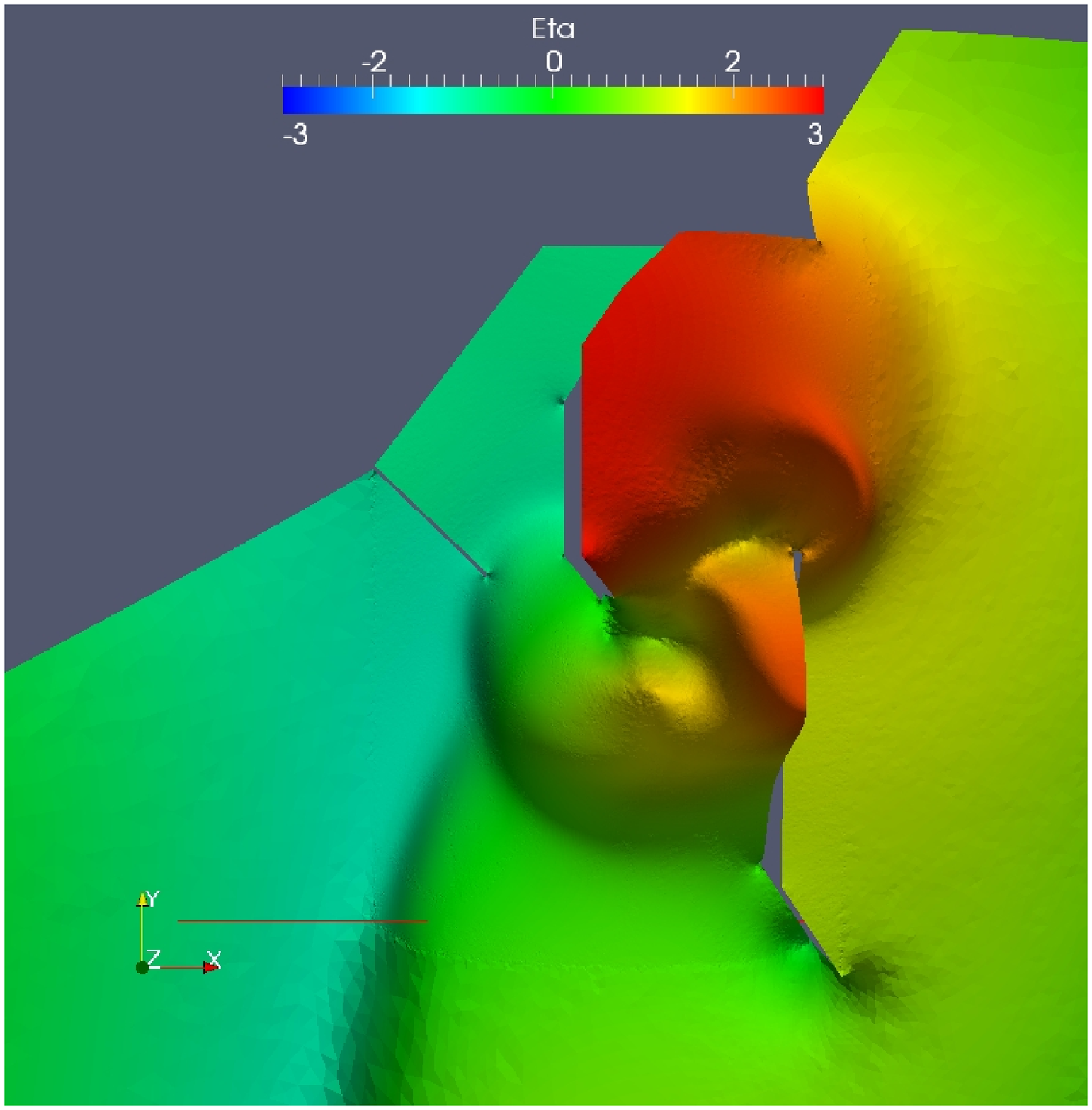} 
}
\subfigure[$t = 14$ min]{
\includegraphics[width=0.31\textwidth]{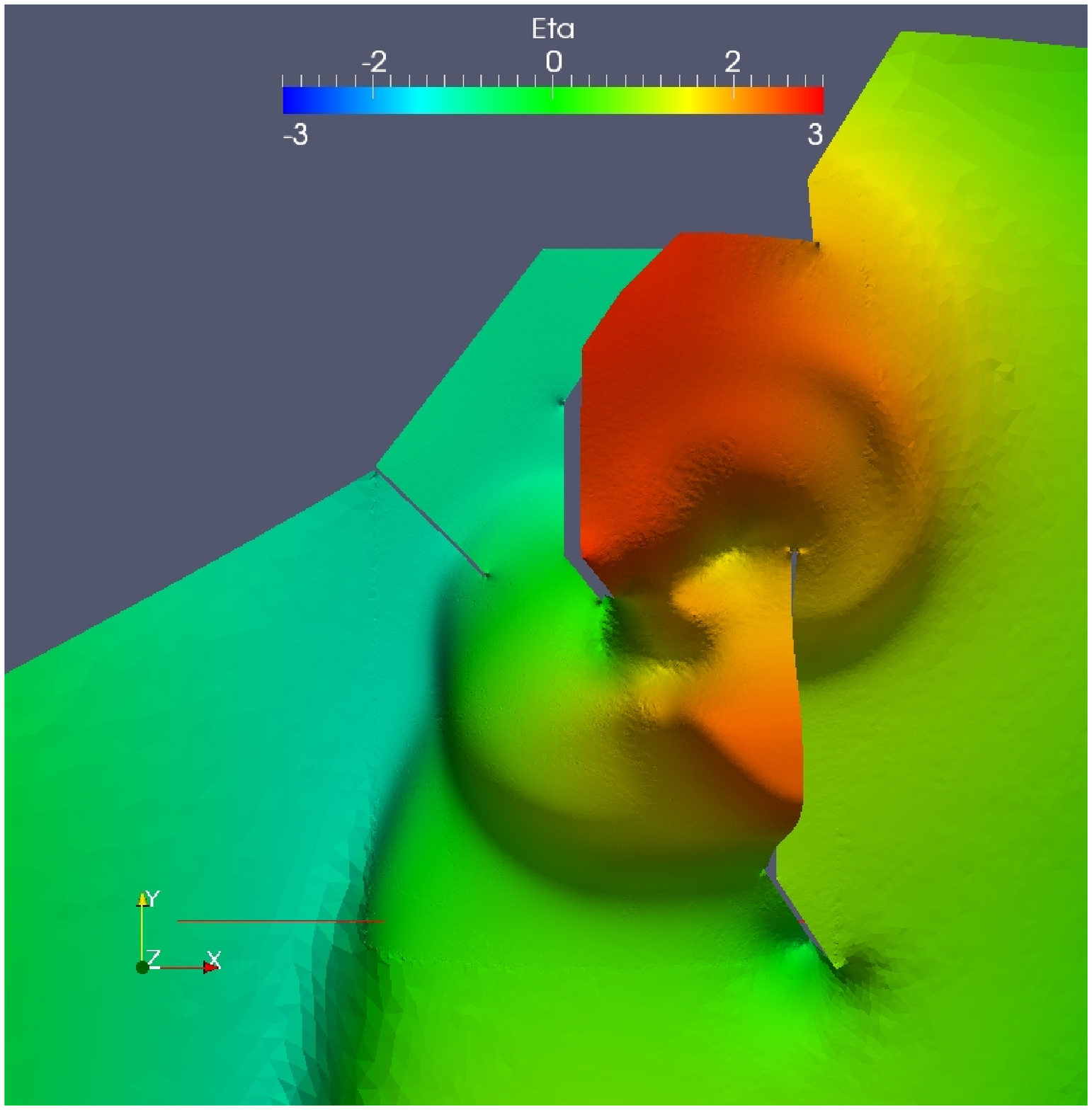} 
}
\subfigure[$t = 14$ min $20$ sec]{
\includegraphics[width=0.31\textwidth]{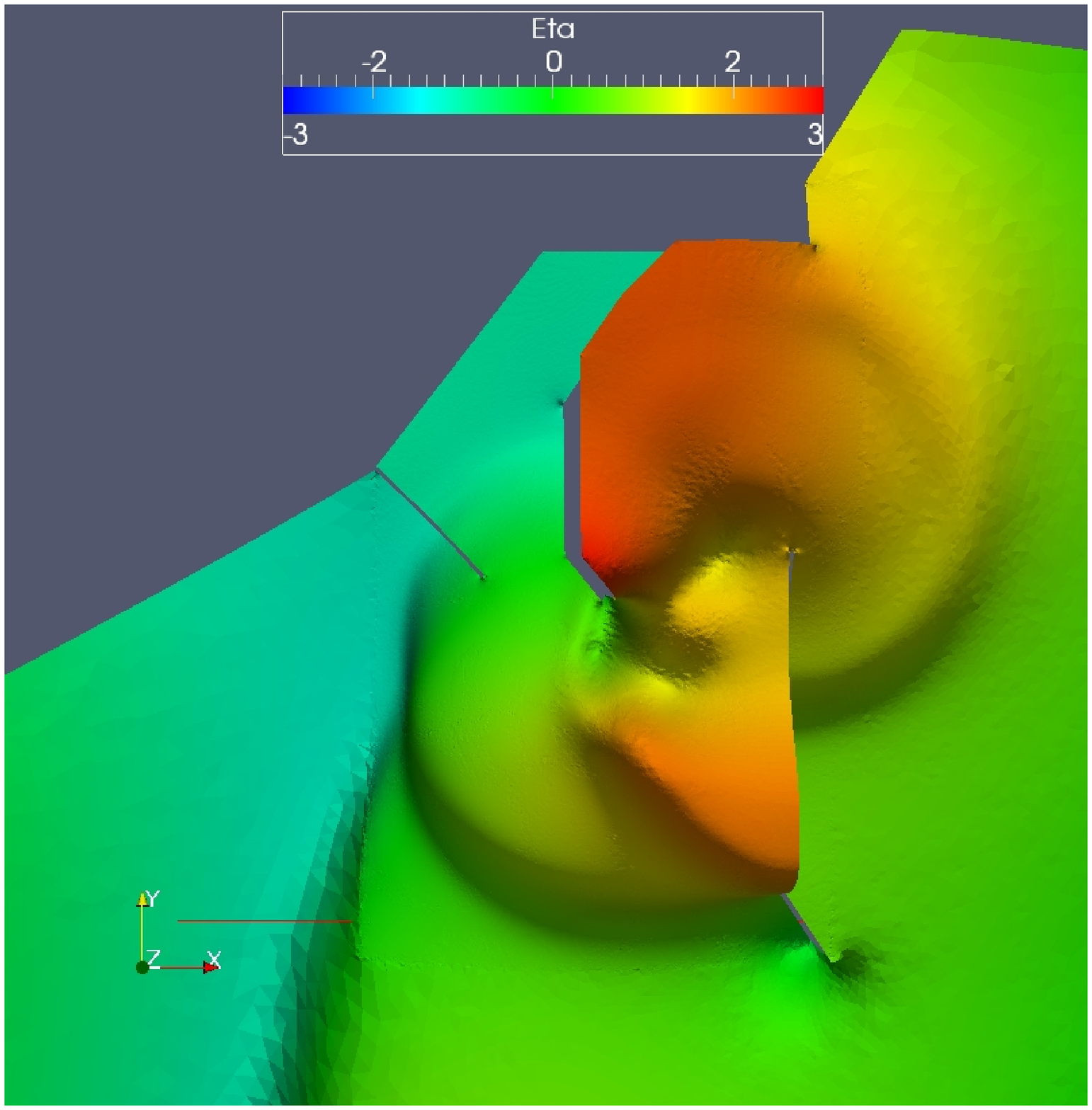} 
}
\subfigure[$t = 14$ min $40$ sec]{
\includegraphics[width=0.31\textwidth]{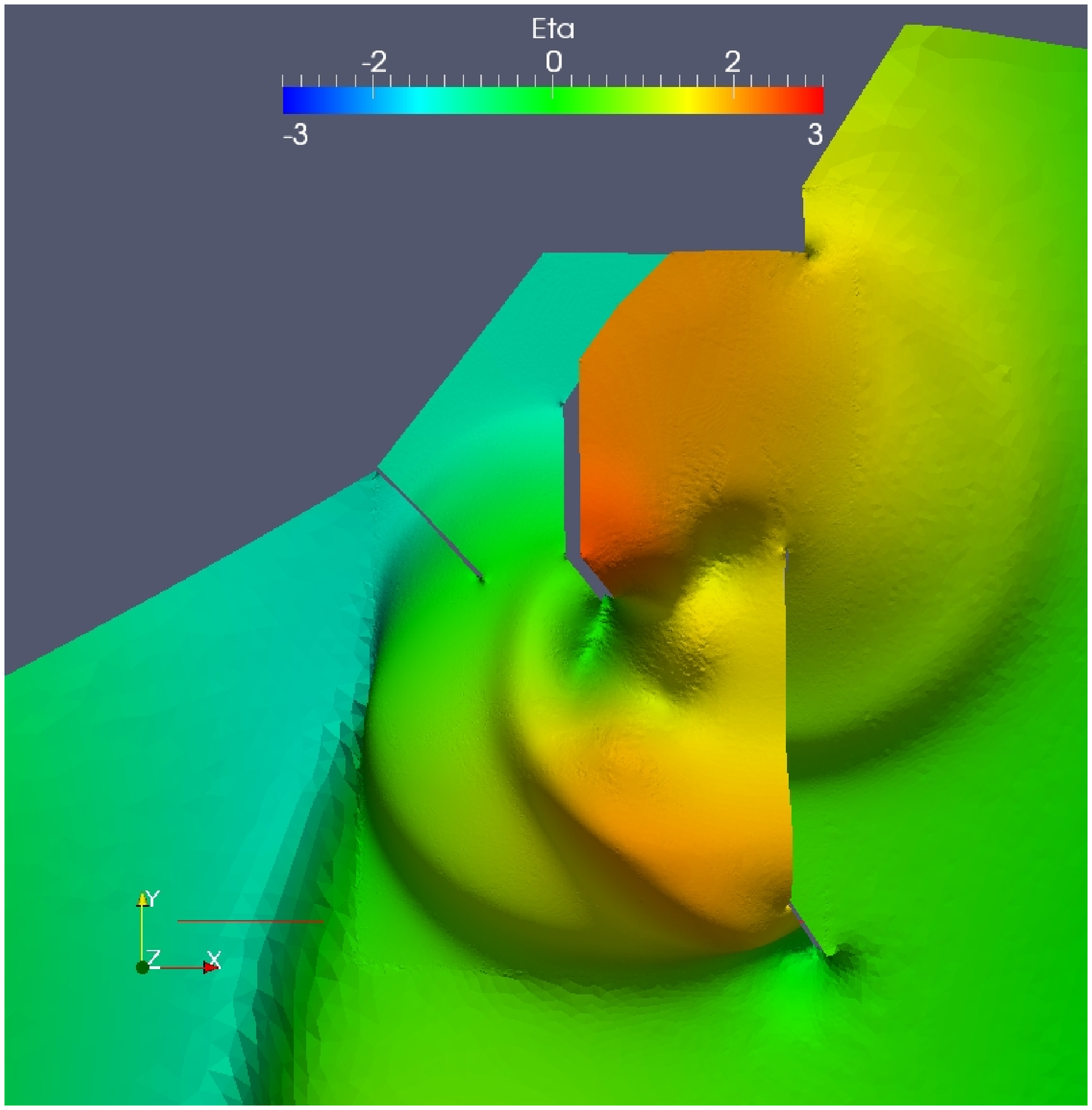} 
}
\subfigure[$t = 15$ min $10$ sec]{
\includegraphics[width=0.31\textwidth]{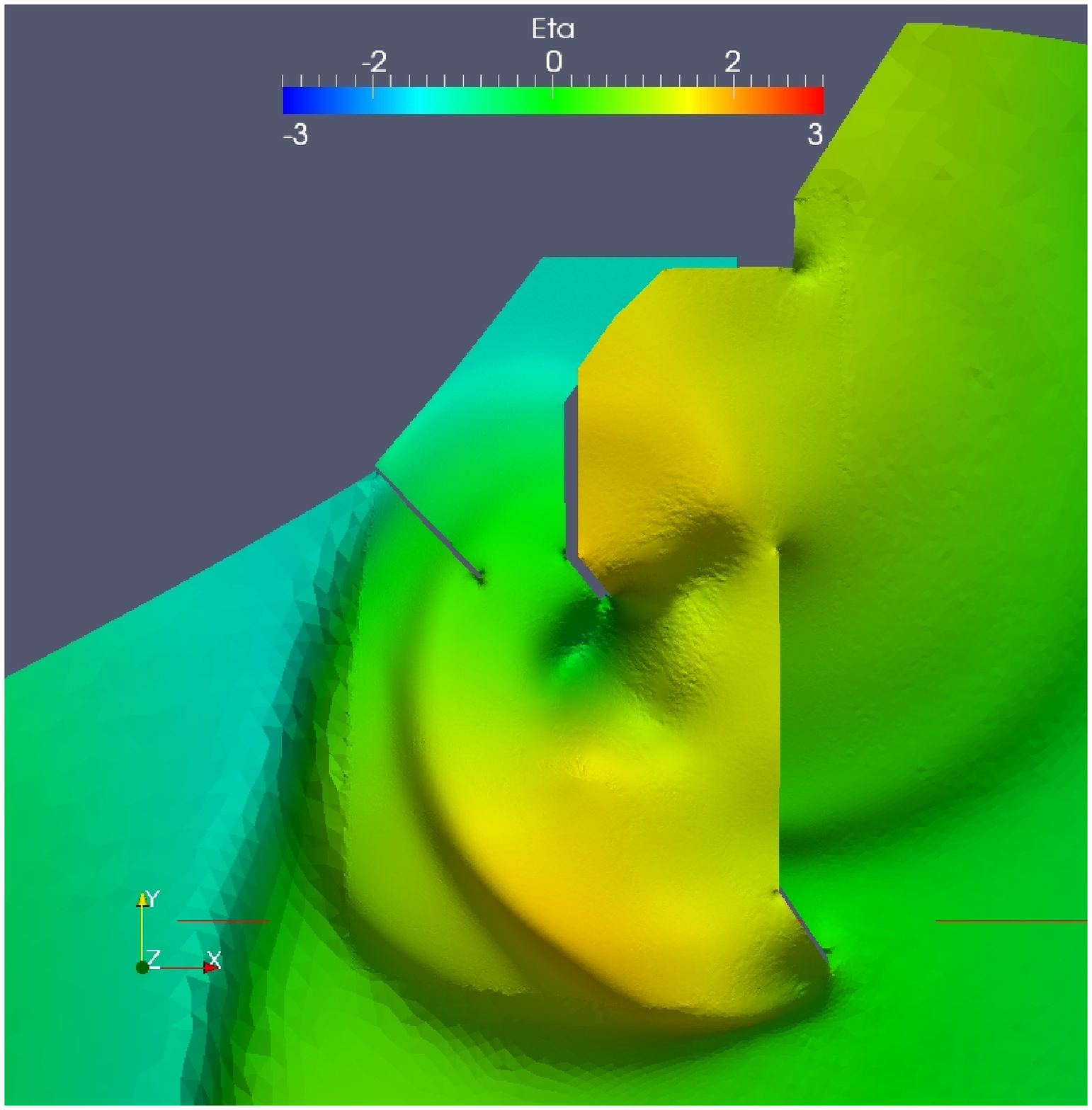} 
}
\subfigure[$t = 15$ min $30$ sec]{
\includegraphics[width=0.31\textwidth]{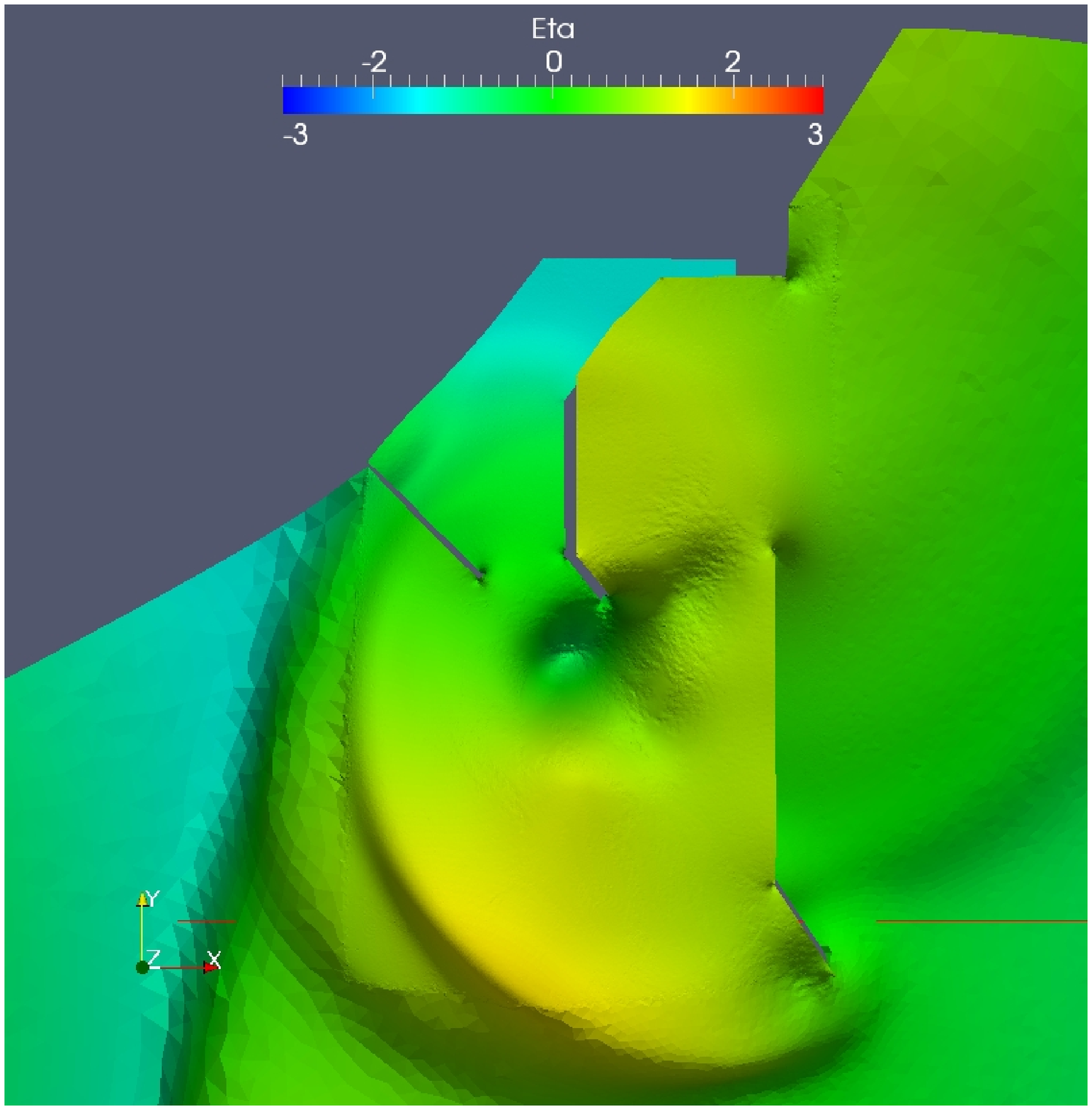} 
}
\subfigure[$t = 15$ min $50$ sec]{
\includegraphics[width=0.31\textwidth]{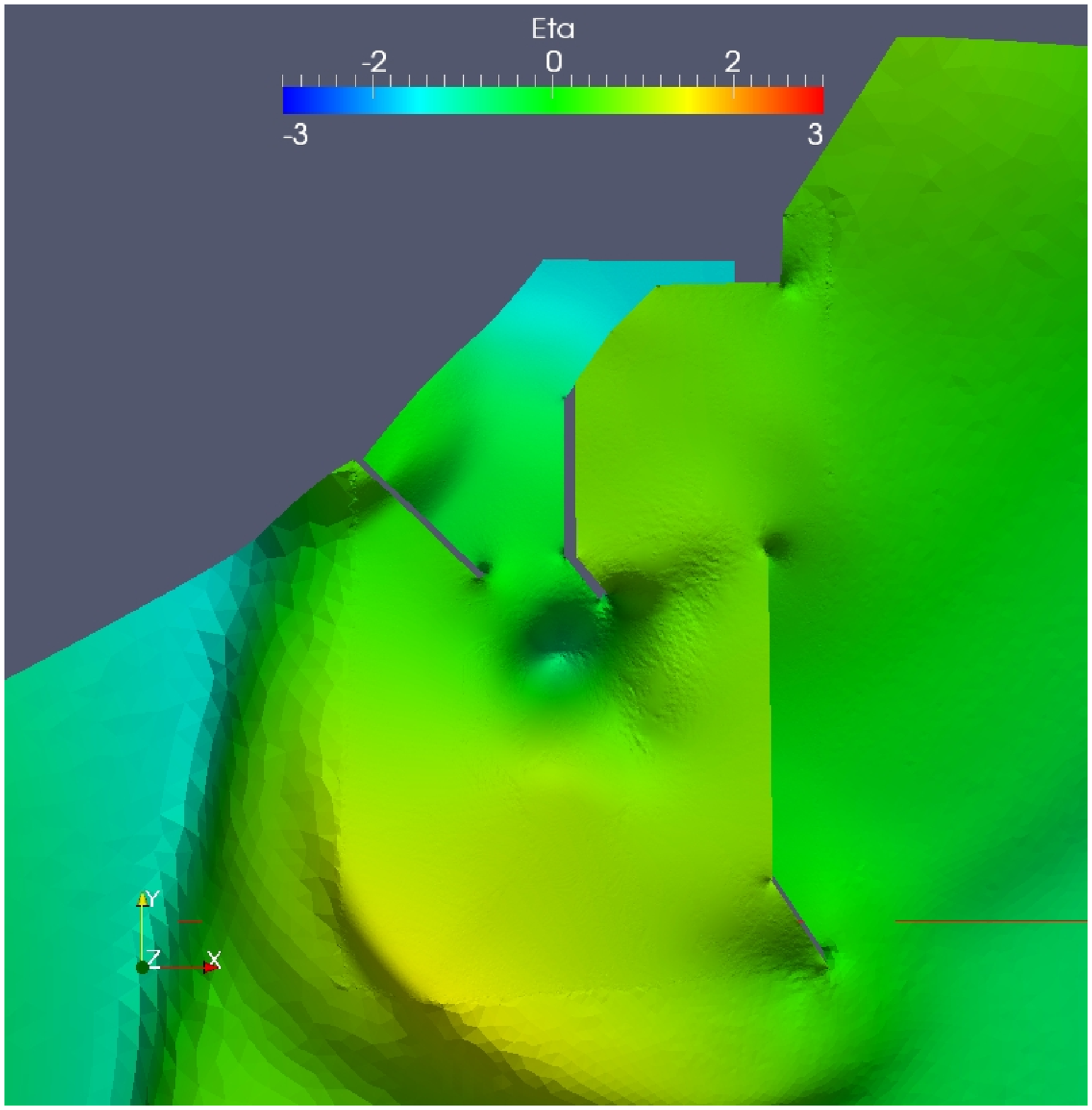} 
}
\end{center}
\caption{Free-surface close up between $13$ and $16$ minutes.}
\label{Oarai_results_close}
\end{figure}

\begin{figure}
\begin{center}
\subfigure[$t = 18$ min]{
\includegraphics[width=0.45\textwidth]{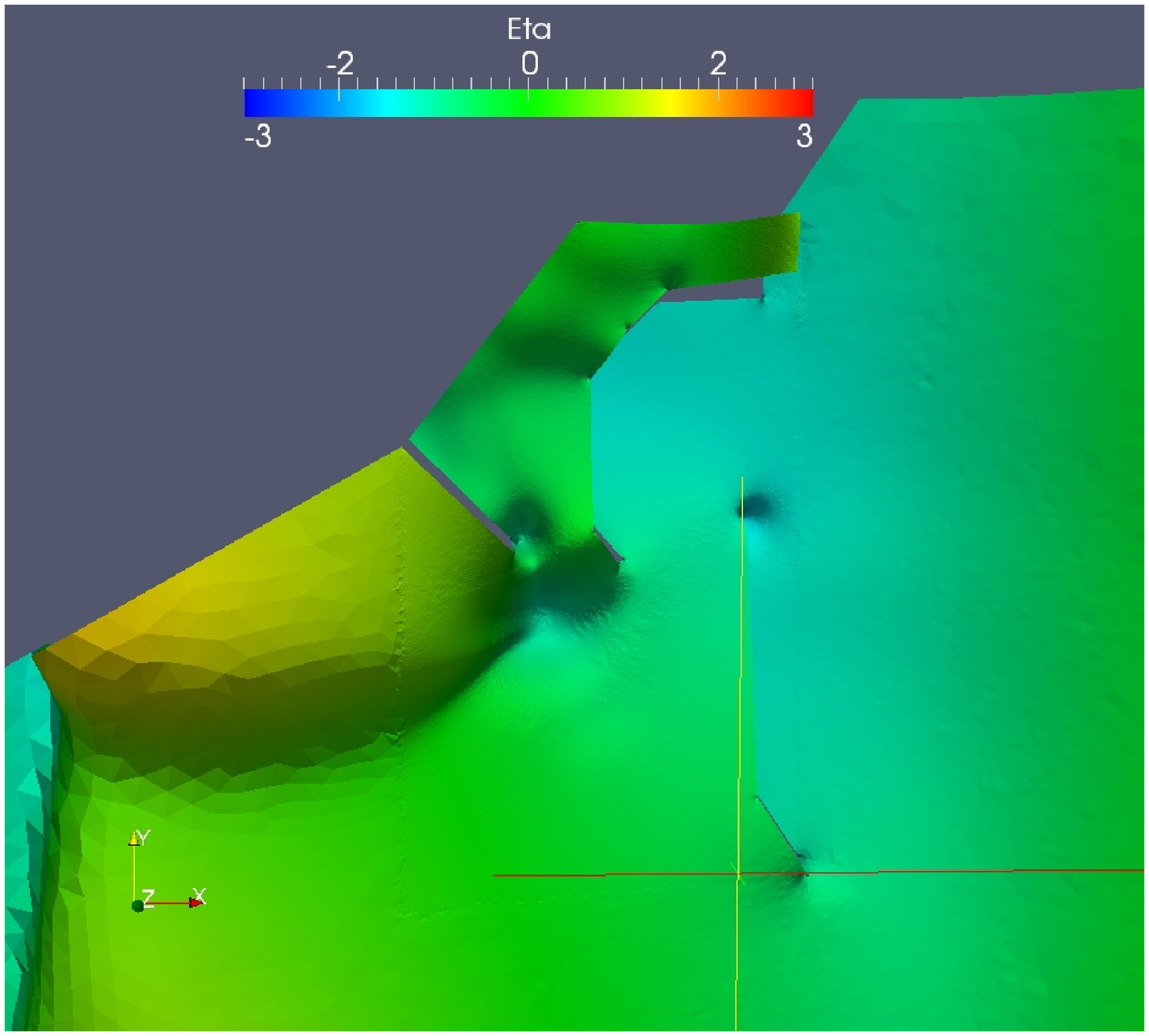}
}
\subfigure[$t = 20$ min]{
\includegraphics[width=0.45\textwidth]{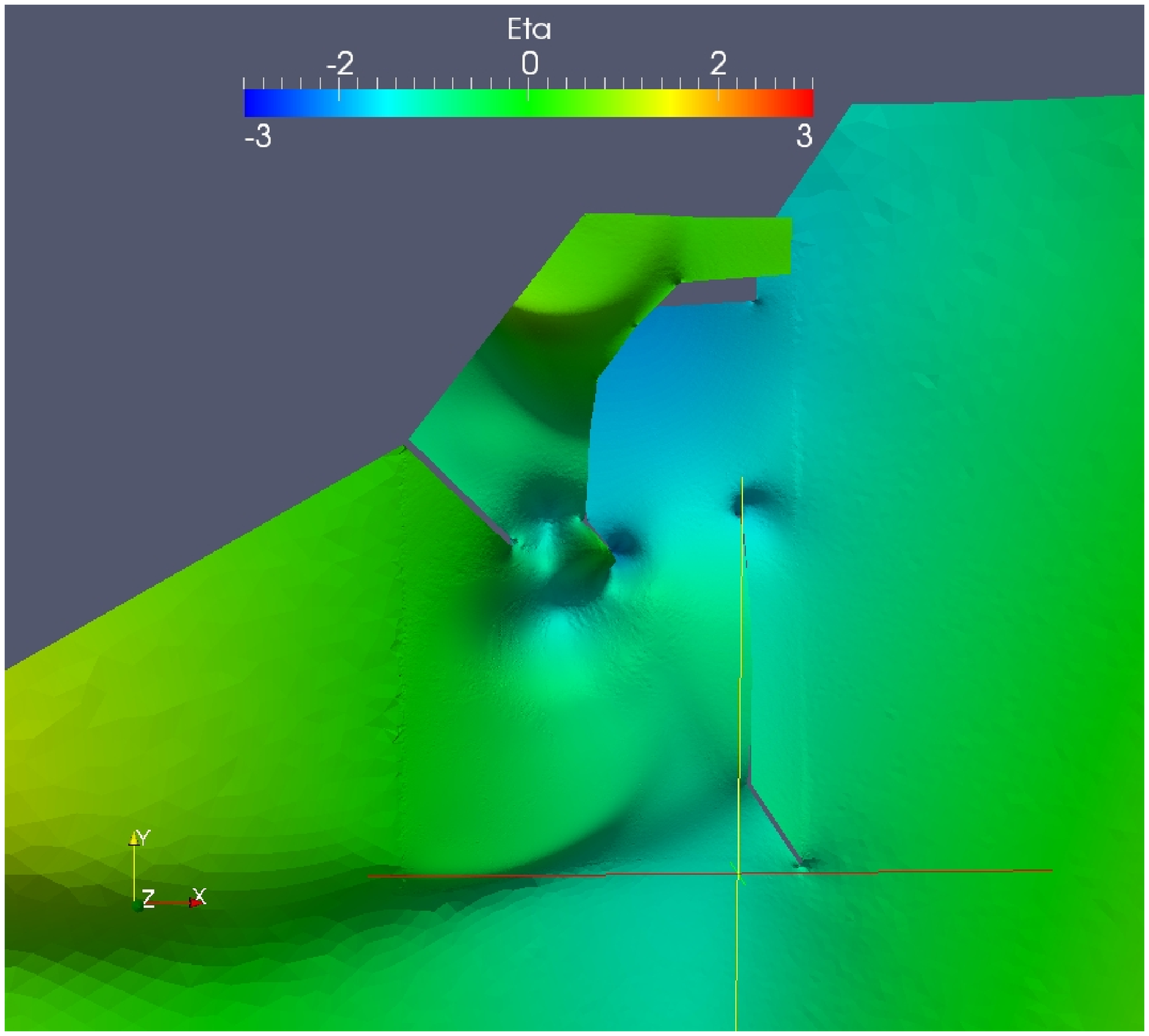} 
}
\subfigure[$t = 24$ min]{
\includegraphics[width=0.45\textwidth]{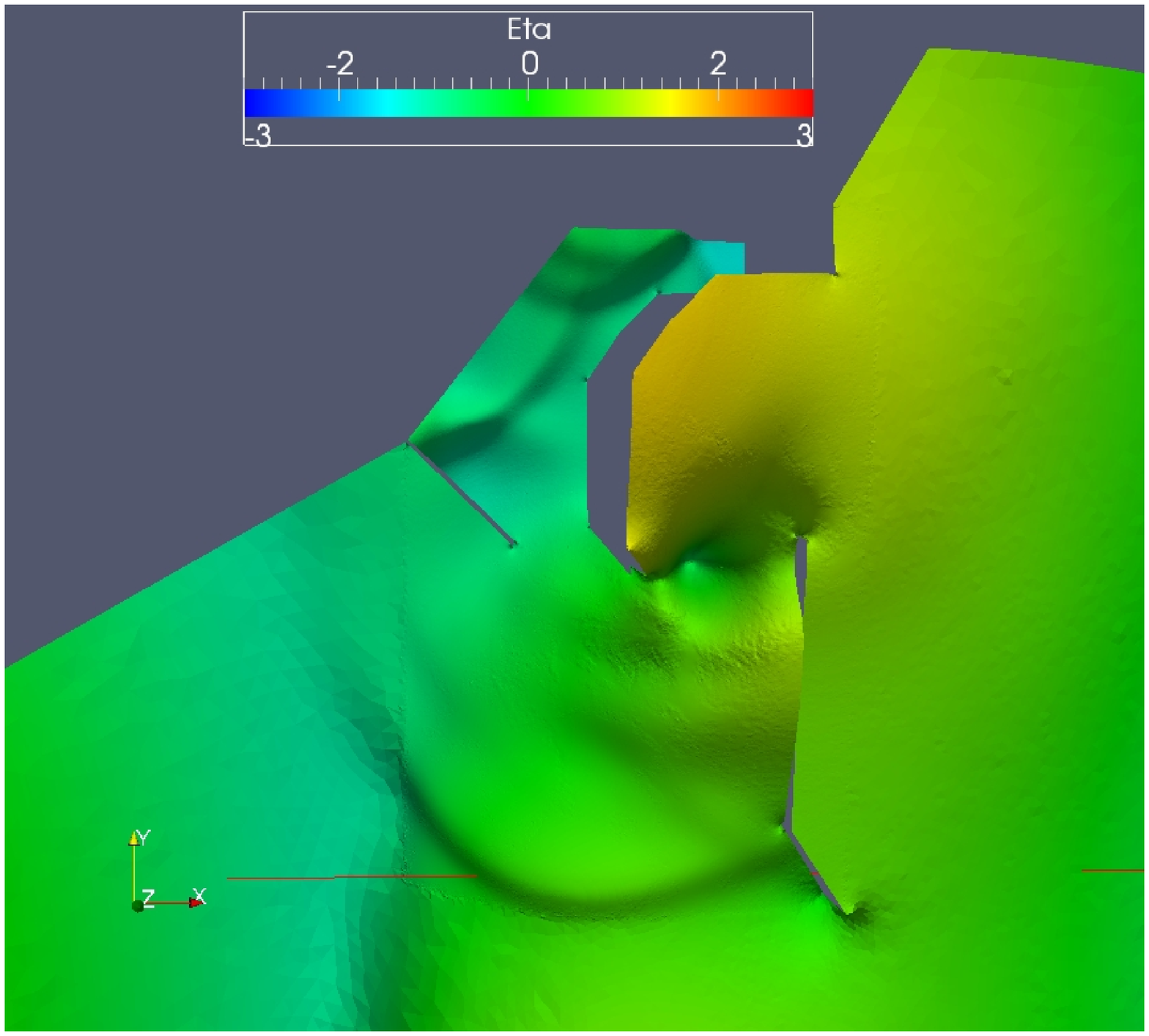} 
}
\subfigure[$t = 28$ min]{
\includegraphics[width=0.45\textwidth]{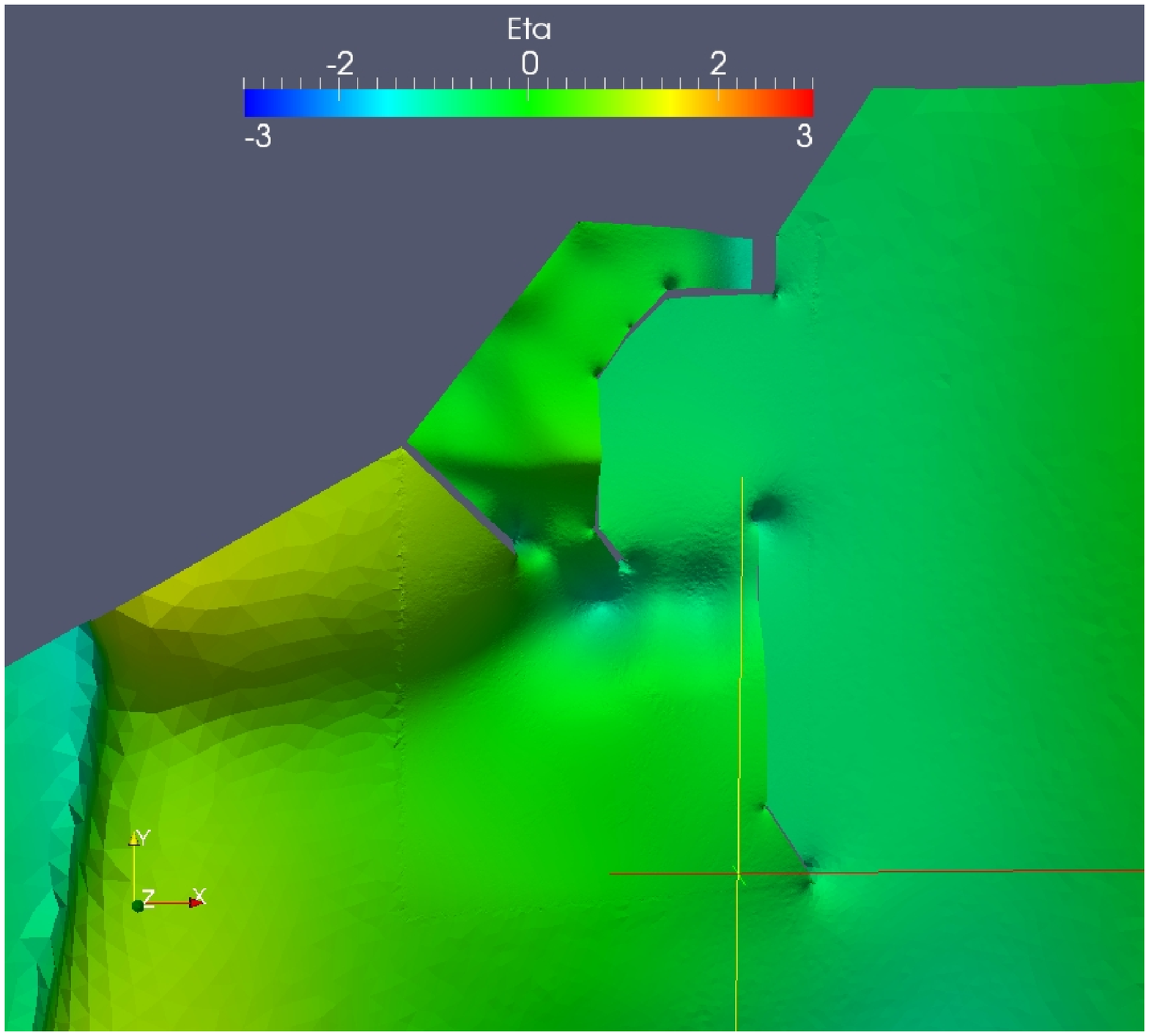} 
}
\end{center}
\caption{Free-surface close up at later times.}
\label{Oarai_results_close_later}
\end{figure}

The free surface results between $7$ and $18$ minutes are shown in Figure \ref{Oarai_results}. Once the wave hits the first harbour wall, it is reflected and refracted. When the wave hits the subsequent boundaries a swirling type motion starts to appear (see Figure \ref{Oarai_results_close} for a close up of this effect between $13$ and $16$ minutes). This motion is dramatic for a few minutes, however even sometime later a depression in the harbour is maintained (see Figure \ref{Oarai_results_close_later}). It is important to point out that the observed vortices are macro-scale ones since the motion is assumed to be non-dissipative. More details can be found in \cite{O'Brien2012}.

\section{Conclusion}
This paper shows that there are still a number of poorly explored topics related to tsunami science, especially when it comes to tsunami generation or to tsunami run-up. 

\section*{Acknowledgements}
This work was funded by IRCSET under the Postgraduate Research Scholarship Scheme, ERC under the research project ERC-2011-AdG 290562-MULTIWAVE and SFI under the programme ERC Starter Grant -- Top Up, Grant 12/ERC/E2227.

\bibliography{biblio}
\bibliographystyle{plain}

\end{document}